\documentclass[preprint,showpacs,titlepage,aps,prd,
tightenlines,
amsmath,byrevtex,nofootinbib]{revtex4}

\usepackage{graphicx}

\begin{document}

\preprint{hep-ph/0505133}

\title{Testing CPT Symmetry with Supernova Neutrinos}

%\vskip 2cm

\author{Hisakazu Minakata}
\email{E-mail: minakata@phys.metro-u.ac.jp}
\author{Shoichi Uchinami}
\email{E-mail: uchinami@phys.metro-u.ac.jp}
\affiliation{Department of Physics, Tokyo Metropolitan University \\
1-1 Minami-Osawa, Hachioji, Tokyo 192-0397, Japan}

\date{August 13, 2005}

\vglue 1.6cm
%%%%%%%%%%%%%%%%%%%%%%%%%%%%%%%%%%%%%%%%%%%%%%%
%    Abstract
%%%%%%%%%%%%%%%%%%%%%%%%%%%%%%%%%%%%%%%%%%%%%%%

\begin{abstract}

Diagnosing core of supernova requires favor-dependent 
reconstruction of three species of neutrino spectra, 
$\nu_e$, $\bar{\nu}_{e}$ and $\nu_x$ 
(a collective notation for $\nu_{\mu}$, $\bar{\nu}_{\mu}$, $\nu_{\tau}$, and 
$\bar{\nu}_{\tau}$). 
We point out that, assuming the information available, 
CPT symmetry can be tested with supernova neutrinos. 
We classify all possible level crossing patterns of neutrinos and 
antineutrinos into six cases and show that half of them contains only 
the CPT violating mass and mixing patterns. 
We discuss how additional informations from terrestrial experiments 
help identifying CPT violation by narrowing down the possible flux patterns. 
Although the method may not be good at precision test, 
it is particularly suited to uncover gross violation of CPT such as 
different mass patterns of neutrinos and antineutrinos. 
The power of the method is due to the 
nature of level crossing in supernova which results in the sensitivity 
to neutrino mass hierarchy and to the unique characteristics 
of in situ preparation of both $\nu$ and $\bar{\nu}$ beams. 
Implications of our discussion to the conventional analyses with 
CPT invariance are also briefly mentioned.

\end{abstract}


\maketitle

%%%%%%%%%%%%%%%%%%%%%%%%

\section{Introduction}
\label{introduction}

CPT is one of the most fundamental symmetries in relativistic 
quantum field theory \cite{weinberg}, by which 
masses and flavor mixing angles are constrained to be identical 
for particles and their antiparticles. 
Because of the fundamental nature of the symmetry, it is 
important to test the CPT invariance, and there has been 
continuing efforts mainly in kaon physics \cite{PDG}. 
Needless to say, the effort of testing CPT symmetry should 
be extended to the lepton sector.

Recently, there arose some interests in possible violation of 
CPT symmetry in the lepton sector, partly motivated by 
a possible interpretation of the LSND result \cite{LSND}. 
A hypothesis of different mass and mixing patterns of neutrino 
and antineutrino sectors, as first suggested by 
Murayama and Yanagida \cite{MY00}, 
could be flexible enough to accommodate the LSND data in the 
three-neutrino framework, while not sacrificing 
the success of describing the atmospheric, the solar, the reactor, 
and the accelerator data \cite{SKatm,solar,KamLAND,K2K}.
While the proposal was followed by a series of papers 
\cite{Barenboim,Strumia02,Gouvea02}, 
it was shown by Gonzalez-Garcia, Maltoni and Schwetz \cite{concha03}, 
in an extensive statistical analysis of all the data including 
KamLAND \cite{KamLAND}, that the CPT violating hypothesis 
is not in good shape. 
Interestingly, the best fit point of all data except for LSND is 
CPT symmetric, and the mixing parameter region favored by 
LSND is more than 3$\sigma$ away from the region favored by 
all but LSND data.

In this paper, we explore a possibility of testing CPT symmetry 
with supernova neutrinos. 
Independent of the success or the failure of the CPT violating 
scenario for the LSND data, it is important to test CPT as a 
fundamental symmetry in nature. 
Given the fact that neutrinos has brought us several surprises, 
there exists an even more intriguing (albeit not likely) possibility 
to discover CPT violation by future neutrino experiments.
Supernova neutrinos are advantageous to examine neutrino and 
antineutrino properties simultaneously and consistently 
because the beam is composed not only of 
$\nu_e$,  $\nu_{\mu}$, and $\nu_{\tau}$ but also their antiparticles.
We examine the possibility of using neutrinos from supernova 
to identify CPT violation assuming the resolving power of 
flavor-dependent neutrino fluxes 
in future observation of galactic supernovae.

So far, constraints on CPT violation of mixing parameters 
in the lepton sector have been derived by Super-Kamiokande (SK) 
group \cite{SK_CPT} and in \cite{concha03}. 
Possible ways of testing CPT symmetry has been discussed 
by using solar and reactor neutrinos 
\cite{Bahcall02,MNTZ}, and neutrino factory \cite{Bilenky01}. 
We restrict ourselves, as these preceding works do, to the 
framework of possible CPT violation in masses and flavor mixing 
of neutrinos, assuming that neutrino interactions conserve CPT.
Then, the natural question is how supernova method can be 
competitive to these more ``traditional'' methods for testing CPT. 
It is the right question because it is unlikely that such rare event 
as supernova can be used for a precision test of CPT symmetry. 
Despite the reasonable skepticism, we will show 
in this paper that supernova neutrino can be a powerful tool for 
uncovering gross violation of CPT symmetry. 
(See Sec.~\ref{whyhow} for further comments.)

We rely on the reference \cite{dighe-smi} for the formulas of  
neutrino flavor conversion in supernova in the three-flavor framework. 
We should note that our general CPT non-invariant treatment, 
of course, includes the case of CPT invariance. 
Therefore, the reader can use part of the formulas given in this 
paper as a compact recollection of those in \cite{dighe-smi}, 
but by now without ambiguities due to the solar neutrino solutions.

In Sec.~\ref{whyhow}, we discuss the question of why and how 
supernova neutrinos are useful to test CPT. 
In Sec.~\ref{review}, we review the basic properties of supernova 
neutrinos and the approximations involved in our treatment. 
In Sec.~\ref{classify}, 
after recollecting compact formulas for neutrino flavor 
conversion in supernova, we present a complete classification 
of spectral patterns of supernova neutrinos that are 
possible in a general CPT violating ansatz. 
In Sec.~\ref{reduced}, 
we discuss how the allowed flux patterns of supernova neutrinos reduce 
as additional input of $\theta_{13}$ and neutrino mass hierarchy are added. 
In Sec.~\ref{characteristic}, we give a comparative study of 
characteristic features of spectra of three effective neutrino species 
predicted in each classified patterns of 
neutrino flavor transformation in supernova. 
In Sec.~\ref{howwell}, 
we  discuss at a qualitative (or semi-quantitative) level to what 
extent the possible different 
neutrino flux patterns can be discriminated observationally.
In Sec.~\ref{conclusion}, we give the concluding remarks.

\section{Why and how are supernova neutrinos useful to test CPT?}
\label{whyhow}

To answer the question of why and how supernova neutrinos are 
useful to test CPT we need to specify the question; Namely, 
which aspects of CPT symmetry do we want to test, or which features 
of CPT violation do we try to uncover?

First of all, lacking well defined models of CPT violation, we cannot 
test it in neutrino interactions by using supernova neutrinos. 
If the interactions are different from what we know the neutrino 
properties inside the core must be recomputed with new interactions 
to define CPT non-invariant features of supernova neutrinos. 
We do not have the recipe to carry it out. 
Therefore, we restrict the type of CPT violation to test 
to the ones signaled by difference between masses and mixing parameters 
of neutrinos and antineutrinos, assuming that their interactions 
are described by the standard model.

How can CPT violation be actually signaled by supernova neutrinos? 
As we will show in the subsequent sections, possible difference in 
mass patterns and mixing angles of neutrinos and antineutrinos 
results in several different spectral patterns of three species of 
neutrinos, $\nu_e$, $\bar{\nu}_{e}$, and $\nu_x$, 
where the last is a collective notation 
for $\nu_{\mu}$, $\bar{\nu}_{\mu}$, $\nu_{\tau}$, and 
$\bar{\nu}_{\tau}$. (See Sec.~\ref{review}.)
In this paper, we rely on the assumption that flavor-dependent 
reconstruction of supernova neutrino fluxes will be done at the time of 
next supernova, so that 
CPT violating patterns of neutrino spectra can be identified. 
It may be realized either by arrays of detectors of various types, 
or by a limited number of them with some ingenious method for analysis.  
Though highly nontrivial, this type of flavor-dependent 
reconstruction of supernova neutrino spectra is required anyway 
to diagnose core of the supernova in the conventional analysis 
assuming CPT invariance.
The importance of the last point has been emphasized since 
sometime ago \cite{diagnostics}.

Although we formulate the problem of testing CPT with supernova 
neutrinos in a generic way, we focus on the gross (or ``discrete'') 
violation of CPT caused by different mass patterns and/or by 
possible large difference in mixing angle $\theta_{13}$ in 
neutrinos and antineutrinos sectors.
Then, we show that supernova neutrino can be a 
powerful indicator of CPT violation.  
Signaling CPT violation can be done by distinguishing 
spectral patterns of three effective species of neutrinos 
characteristic to CPT violation, which come from unequal level 
crossing patterns at the high-density resonance 
of neutrinos and antineutrinos. 
It should also be noted that this type of CPT test is quite 
complementary to the one which measures small differences of 
neutrino and antineutrino mixing parameters. 
The method of looking for gross violation of CPT is  
quite insensitive to the presence of tiny difference 
in mixing parameters because the effect we are looking for is 
robust and only depends upon mass patterns.  
The issue of supernova neutrino as a sensitive probe for neutrino mass 
hierarchy was first discussed in \cite{MN_inverted}.\footnote{
%%%%%%%%%%%%%%%% footnote %%%%%%%%%%%%%%%%%
The paper contains, in addition to the general statement of utility of 
supernova neutrinos as a tool of discriminating mass hierarchy, 
an analysis of SN1987A data which leads the authors to conclude that 
(in page 306) ``if the temperature ratio $\tau \equiv T_{\nu_x} / T_{\bar{\nu}_{e}}$ 
is in the range 1.4-2.0 as the SN simulations indicate, the inverted hierarchy 
of neutrino masses is disfavored by the neutrino data of SN1987A 
unless the H resonance is nonadiabatic''. 
While it follows the spirit of the earlier analyses 
\cite{SSB94,jegerlehner}, our analysis using the 
three-flavor mixing framework has physics consequences quite 
different from the ones spelled out in these papers.  
In fact, the ansatz tested in \cite{jegerlehner} 
is different from the hypothesis we have tested 
(which was relatively more disfavored) 
due to the three-flavor treatment of the problem. 
We note that most of the criticism posed by Barger {\it et al.} \cite{barger02} 
does {\it not} apply to our analysis because it does not rely on 
the goodness of fit in the likelihood analysis 
but on the credibility of the parameters obtained as a result of fit. 
The nature of this type of analysis was already made fully transparent   
by Jegerlehner, Neubig, and Raffelt \cite{jegerlehner} in their thorough 
analysis done in 1996 \cite{raffelt}.
It would be very interesting to come back to the debate after having 
supernova simulations calibrated by the high-statistics data of future 
galactic supernova. 
}

\section{Basic properties of supernova neutrinos and 
approximations involved in the treatment}
\label{review}

In this section, we briefly summarize the basic properties 
of supernova neutrinos and the approximations involved in 
our treatment. Our description will be a very brief one and 
we refer \cite{dighe-smi} for detailed discussions, on which 
our treatment and notations will be based. 
The great simplification that occurred after the work is 
published is that the large mixing angle (LMA) region   
of the solar Mikheyev-Smirnov-Wolfenstein (MSW) solution \cite{MSW} is  
selected out both in neutrino and antineutrino sectors by 
all the solar and the KamLAND experiments, respectively 
\cite{solar,KamLAND}. 

We assume three flavor mixing scheme of 
neutrinos with the standard form~\cite{PDG} of lepton flavor 
mixing matrix, the Maki-Nakagawa-Sakata (MNS) matrix~\cite{MNS},
\begin{eqnarray}
U=\left[
\begin{array}{ccc}
c_{12}c_{13} & s_{12}c_{13} &   s_{13}e^{-i\delta}\\
-s_{12}c_{23}-c_{12}s_{23}s_{13}e^{i\delta} &
c_{12}c_{23}-s_{12}s_{23}s_{13}e^{i\delta} & s_{23}c_{13}\\
s_{12}s_{23}-c_{12}c_{23}s_{13}e^{i\delta} &
-c_{12}s_{23}-s_{12}c_{23}s_{13}e^{i\delta} & c_{23}c_{13}\\
\end{array}
\right],
\label{MNSmatrix}
\end{eqnarray}
where  $c_{ij}$ and $s_{ij}$ ($i, j = 1 \mbox{-} 3$) imply 
$\cos{\theta_{ij}}$ and $\sin{\theta_{ij}}$,  
respectively.
The lepton mixing matrix $U$ relates the flavor eigenstate to the 
mass eigenstate as $\nu_{\alpha} = U_{\alpha i} \nu_i$, where 
$\alpha = e, \mu, \tau$ and $i=1, 2, 3$. 
The mass squared difference of neutrinos 
is defined as $\Delta m^2_{ij} \equiv m^2_i - m^2_j$ where
$m_i$ is the eigenvalue of the $i$th mass-eigenstate.
To distinguish antineutrino mixing matrix from that of neutrinos 
we place a ``bar" onto the corresponding mixing parameters.

In the analysis in this paper, we restrict ourselves into the 
simplified ansatz for supernova neutrinos. That is, 
$\nu_{\mu}$, $\bar{\nu}_{\mu}$, $\nu_{\tau}$, and 
$\bar{\nu}_{\tau}$ are treated as a single component 
denoted as $\nu_x$. It is a good approximation because 
they interact with surrounding matter only through neutral 
current (NC) interactions, and hence they are practically 
physically indistinguishable with each other.
Under the approximation, supernova neutrinos consist of 
the three components, 
$\nu_e$, $\bar{\nu}_{e}$ and $\nu_x$.

It is in fact very simple to compute the neutrino flux just outside 
supernova for a given set of neutrino fluxes at neutrino sphere. 
To do this one first draw the level crossing diagrams of neutrinos 
and antineutrinos, as given in Fig.~\ref{levelcross}. 
The characteristic feature of the supernova neutrino level 
crossing diagram, which is unique among astrophysical objects, 
is that there are two resonances, 
one in high and the other low density regions,  corresponding 
respectively to the atmospheric and the solar $\Delta m^2$ scales. 
(In a typical supernova progenitor, they are located in 
helium and hydrogen burning shells, respectively.)
They are referred as the H and L level crossings in this paper. 
The level crossing pattern as well as their (non-) adiabaticity are the 
decisive factors of neutrino flavor conversion in supernova \cite{MN90}.

It should be noticed that, because we are preparing a general 
CPT non-invariant framework, we have to draw diagrams of 
$\nu$ and $\bar{\nu}$ separately, thereby allowing the cases 
with different mass hierarchies for neutrinos and antineutrinos.
Altogether there are two and four cases of level crossing diagram,  
corresponding to the normal and the inverted mass hierarchies in 
$\nu$ and $\bar{\nu}$ sectors, respectively. 
Proliferation of $\bar{\nu}$ diagram by a factor of 2 is due to 
inability of distinguishing $\bar{m}_2 > \bar{m}_1$ or $\bar{m}_2 < \bar{m}_1$ cases. 
It should be noticed that 
one can adopt one of the two conventions, which are equivalent 
with each other: 
(1) $\bar{m}_2 > \bar{m}_1$ and $0 < \theta < \pi/2$, 
or 
(2) $0 < \theta < \pi/4$ with $\bar{m}_2 > \bar{m}_1$ or $\bar{m}_2 < \bar{m}_1$.
In this paper, we take the latter convention.

An enormous simplification results in the treatment of neutrino 
flavor transformation in supernova 
(in fact in the envelope of the progenitor star) if the two resonances, 
H and L, are approximately independent with each other. 
It was argued in \cite{dighe-smi} that they are, based on a 
factor of $\simeq$30 difference between $\Delta m^2_{31}$ 
and $\Delta m^2_{21}$ but under the assumption that they 
are identical in neutrinos and antineutrino sectors.
%Let us reexamine this point by allowing CPT violation. 
%
Fortunately, thanks to the currently available constraints on 
$\Delta m^2$ and $\bar{\Delta m^2}$ which are already 
rather powerful, we can argue that the same approximation 
applies even when we relax the assumption that they are identical.
We first note that $\Delta m^2_{21}$ and $\bar{\Delta m^2}_{21}$ 
are both in the  ``LMA'' region; 
They are constrained to be in the regions 
\begin{eqnarray}
2 \times 10^{-5} \mbox{eV}^2 \leq  &\Delta m^2_{21}& \leq 2 \times10^{-4} \mbox{eV}^2,
\nonumber \\
10^{-5} \mbox{eV}^2 \leq  &|\bar{\Delta m^2}_{21}| & \leq 2 \times10^{-4} \mbox{eV}^2, 
\label{solarKLbound}
\end{eqnarray} 
former by all the solar neutrino experiments \cite{solar}, 
while the latter by the KamLAND experiment \cite{KamLAND}, 
both at 3 $\sigma$ CL.
On the other hand, $\Delta m^2_{31}$ and $\bar{\Delta m^2}_{31}$ 
are constrained to be 
\begin{eqnarray}
9 \times 10^{-4} \mbox{ eV}^2 \leq &|\Delta m^2_{31}|& \leq 
6 \times 10^{-3} \mbox{ eV}^2, \nonumber \\
4.5 \times 10^{-3} \mbox{ eV}^2 \leq &|\bar{\Delta m^2_{31}}|& \leq 
2 \times 10^{-2} \mbox{ eV}^2, 
\label{SKbound}
\end{eqnarray} 
at 99\% CL by the SK atmospheric neutrino data \cite{SK_CPT}. 
Because of the factor of about 20 difference it can be 
argued quite safely that the approximation of independent 
H and L resonances applies even in our general setting 
which accommodates CPT violation.

It may be appropriate to mention here that the currently available 
bound on possible CPT violation in lepton mixing angles are 
rather mild, as summarized in \cite{MNTZ}. 
The current bound on the difference between
$\sin^2{\theta_{12}}$ for neutrino and
$\sin^2{\bar{\theta}_{12}}$ for antineutrinos
is rather weak \cite{KamLAND}.
Even if we assume that $\bar{\theta}_{12}$ is in the first 
octant,
\begin{eqnarray}
|\sin^2{\theta_{12}} - \sin^2{\bar{\theta}_{12}}| \leq 0.3,
\label{CPT_theta12curr}
\end{eqnarray}
at 99.73\% CL. 
The bound obtained for $\theta_{13}$ is extremely weak, 
$|\sin^2{\theta_{13}} - \sin^2{\bar{\theta}_{13}}|$ can be almost unity 
\cite{concha03}.

\section{Classification of Patterns of Supernova Neutrino Spectra} 
\label{classify}

In this section, after recollecting compact formulas for neutrino flavor 
conversion in supernova (Sec.~\ref{LMA}), we present a complete 
classification of spectral patterns of supernova neutrinos 
(Sec.~\ref{classification}). 
The cases of reduced degeneracy by the aid of additional 
informations from accelerator and reactor experiments will 
be discussed in Sec.~\ref{reduced}. 
The general characteristics of the different flux patterns will be 
described in Sec.~\ref{characteristic}. 
We will discuss in Sec.~\ref{howwell} to what extent these additional 
informations help to discriminate flux patterns by limiting the number 
of possibilities.

\subsection{Neutrino and antineutrino spectra with LMA solution}
\label{LMA}

\begin{figure}[htbp]
%\vglue 0.2cm
\begin{center}
\includegraphics[width=0.9\textwidth]{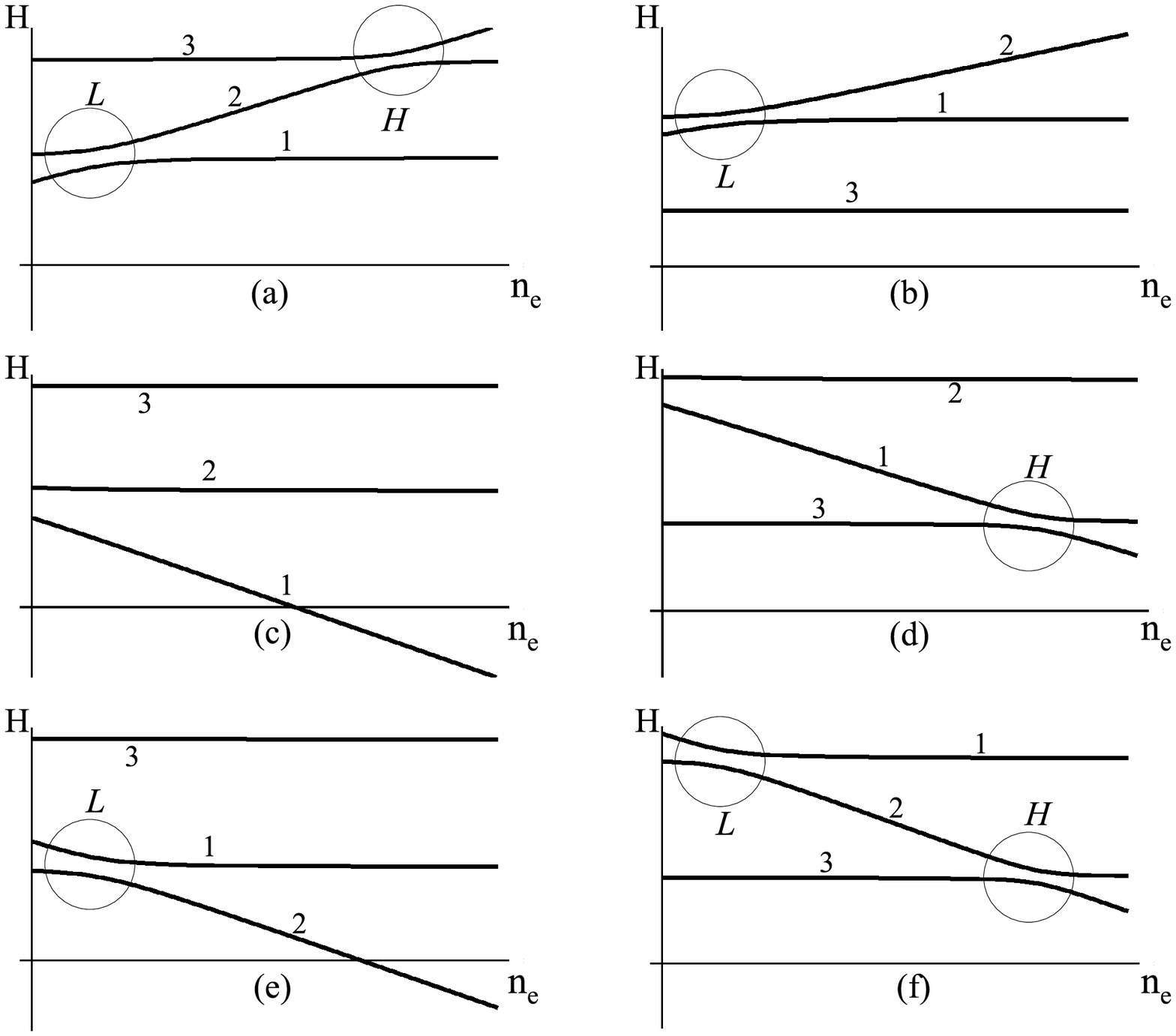}
\end{center}
%\vglue -0.3cm
\caption{Level crossing diagrams for neutrino flavor conversion 
in supernova. In neutrino sector there are 2 diagrams which 
correspond to the normal (a; $\Delta m^2_{31} > 0$) and the 
inverted (b; $\Delta m^2_{31} < 0$) hierarchies. 
In antineutrino sector there are 4 diagrams which 
correspond to the normal  (c, e; $\Delta m^2_{31} > 0$) 
and the inverted  (d, f; $\Delta m^2_{31} < 0$) hierarchies, 
each doubled by two patterns of small mass splittings, 
the reactor-normal (c and d; $\Delta \bar{m}^2_{21}> 0$) 
the reactor-inverted (e and f; $\Delta \bar{m}^2_{21} < 0$) hierarchies.
In CPT invariant case the diagrams (a) to (d) remain. 
}
\label{levelcross}
\end{figure}

Under the approximations spelled out in the previous section, 
the neutrino fluxes that are to reach terrestrial detectors 
can be written in the compact notation \cite{dighe-smi}
\begin{eqnarray}
\left( \begin{array}{c}
F_e \\ F_{\bar{e}} \\ 4 F_x
\end{array} \right) =
\left( \begin{array}{ccc}
p & 0 & 1-p \\ 0 & \bar{p} & 1 - \bar{p} \\
1-p & 1 - \bar{p} & 2 + p + \bar{p}
\end{array} \right) 
\left( \begin{array}{c}
F_e^0 \\ F_{\bar{e}}^0 \\  F_x^0
\end{array} \right)~~.
\label{tr-matrix}
\end{eqnarray}
The above general expression holds for both the normal and 
the inverted mass hierarchies.
By ``normal' and  ``inverted'' we mean 
$\Delta m^2_{31} > 0$ and $\Delta m^2_{31} < 0$, 
which will be denoted hereafter with subscripts $N$ and $I$, respectively, 
Though it is known that $\Delta m^2_{21} > 0$ by the solar neutrino 
observation, there are two possible subclasses in the antineutrino 
sector; 
$\Delta \bar{m}^2_{21} > 0$ and $\Delta \bar{m}^2_{21} < 0$ 
as noted in \cite{gouvea}. 
The former and the latter will be referred to as the 
reactor-normal and the reactor-inverted hierarchies, respectively. 
We will keep this distinction with use of the combined subscript as 
$N21$ and $I21$ ($N12$ and $I12$) for 
$\Delta \bar{m}^2_{21} > 0$ ($\Delta \bar{m}^2_{21} < 0$)  in the case of 
normal and inverted hierarchies, respectively.
Altogether there are two and four different level crossing patterns 
in neutrino and antineutrino sectors, respectively. 
They are depicted in Fig.~\ref{levelcross}.

%%%%%%%%%%%% FIGURE I %%%%%%%%%%%%%%%
%%%%%%%%%%%% FIGURE I %%%%%%%%%%%%%%%

The $\nu_e$ and $\bar{\nu}_e$
survival probabilities $p$ and $\bar{p}$'s are given as 
\begin{eqnarray}
p_{N} &=& |U_{e1}|^2  P_{H} P_{L} + |U_{e2}|^2 P_{H} (1- P_{L})
+ |U_{e3}|^2 (1 - P_{H}) \\
\bar{p}_{N21} &=& |\bar{U}_{e1}|^2  \\
\bar{p}_{N12} &=&  |\bar{U}_{e1}|^2 \bar{P}_L  + |\bar{U}_{e2}|^2 (1 - \bar{P}_L)
\label{p_normal}
\end{eqnarray}
for the normal hierarchy and 

\begin{eqnarray}
p_{I} &=&  |U_{e1}|^2 P_L +  |U_{e2}|^2 (1 - P_L)~~.\\
\bar{p}_{I21} &=&  |\bar{U}_{e1}|^2 \bar{P}_H  +  |\bar{U}_{e3}|^2 (1 - \bar{P}_H) \\
\bar{p}_{I12} &=& |\bar{U}_{e1}|^2 \bar{P}_H \bar{P}_L + 
|\bar{U}_{e2}|^2 \bar{P}_H (1 - \bar{P}_L) +  |\bar{U}_{e3}|^2 (1 - \bar{P}_H) 
\label{p_inverted}
\end{eqnarray}
for the inverted hierarchy.

We note that $\Delta m^2_{21}$ and $\bar{\Delta m^2}_{21}$ 
are both in the  ``LMA'' region; 
The former is constrained to be in the region
$2 \times 10^{-5} \mbox{eV}^2 <  \Delta m^2_{21} <  2 \times10^{-4} \mbox{eV}^2$ 
by all the solar neutrino experiments
while the latter is 
$10^{-5} \mbox{eV}^2 <  \bar{\Delta m^2}_{21} <  2 \times10^{-4} \mbox{eV}^2$ 
by the KamLAND experiments, both at 3 $\sigma$ CL.
Then, one can argue quite safely that the L level crossings 
are adiabatic not only in the neutrino but also in the antineutrino 
channels, $P_L = \bar{P}_L=0$. 
%Although $\bar{\theta}_{12}$ can be in the ``dark side" it does not harm the adiabaticity of the L level crossing. 
%
Then, the survival factors take simple forms 
\begin{eqnarray}
p_{N} &=&  |U_{e2}|^2 P_{H}
+ |U_{e3}|^2 (1 - P_{H}) \\
\bar{p}_{N21}&=& |\bar{U}_{e1}|^2 \\
\bar{p}_{N12}&=& |\bar{U}_{e2}|^2 
\label{p_normal2}
\end{eqnarray}
for the normal hierarchy and 
\begin{eqnarray}
p_{I} &=&  |U_{e2}|^2 \\
\bar{p}_{I21} &=&  |\bar{U}_{e1}|^2 \bar{P}_H + |\bar{U}_{e3}|^2 (1 - \bar{P}_H)\\
\bar{p}_{I12} &=&  |\bar{U}_{e2}|^2 \bar{P}_H + |\bar{U}_{e3}|^2 (1 - \bar{P}_H)
\label{p_inverted2}
\end{eqnarray}
for the inverted hierarchy. 
Therefore, the distinction between 
$\Delta \bar{m}^2_{21} > 0$ and $\Delta \bar{m}^2_{21} < 0$ cases 
is just interchanging $|\bar{U}_{e1}|$ and $|\bar{U}_{e2}|$, as expected.

We make a short remarks on possible roles played by the 
earth matter effect \cite{dighe-smi}. 
It is well known that if the H resonance is adiabatic it plays no role. 
In the case of non-adiabatic H resonance, it can play a role 
but again it is suppressed by the factor 
$(2 E a_{\text{earth}} / \Delta m^2_{31}) \sin^2 2\theta_{13}$, 
where $a = \sqrt{2} G_F N_e(x)$ is related with neutrino's index of
refraction in matter with $G_F$ and $N_e$ being the Fermi constant and 
the electron number density, respectively.
An explicit computation reveals that the effects of the earth matter 
effect is small, and moreover it cannot lift the degeneracy between 
the CPT-conserving and CPT-violating cases.
Therefore, we do not discuss it further in the present paper.

\subsection{General Classification of Patterns of Supernova Neutrino Spectra} 
\label{classification}

In the rest of this paper, we focus on the cases in which the H level 
crossing is completely adiabatic or non-adiabatic; 
We do not discuss the intermediate case in which the H level 
crossing is admixture of adiabatic and non-adiabatic transitions. 
By doing so we restrict ourselves to the two regions of $\theta_{13}$,
roughly speaking, 
$s^2_{13} \geq 10^{-4}$ (adiabatic), or 
$s^2_{13} \leq 10^{-6}$ (non-adiabatic).  
We can classify the possible situation into $4 \times 8 = 32$ 
cases depending upon

\begin{itemize}

\item

neutrino and antineutrino mass patterns are either 
normal or inverted, and if 
$\Delta \bar{m}^2_{21} > 0$ or $\Delta \bar{m}^2_{21} < 0$ 
in the antineutrino sector 

\item

the H level crossings in neutrino and antineutrino sectors 
are adiabatic or non-adiabatic, 

\end{itemize}

\noindent
as given in Table \ref{ad-nadH}. 
In each case, the neutrino fluxes 
$F_e$, $F_{\bar{e}}$, and $F_x$ at a detector can be predicted 
for a given set of $F_e^0$, $F_{\bar{e}}^0$, and $F_x^0$ at 
neutrino sphere.

%%%%%%%%%%%%%%%%% Table I %%%%%%%%%%%%%%%%%%%%
\begin{table}
\vglue 0.5cm
\begin{tabular}{|c|c|c|c|c|c|c|}
\hline
$\nu$/$\bar{\nu}$ survival factor & \hspace{0.55cm} $p_{N}$ \hspace{0.55cm} & \hspace{0.55cm} $p_{I}$  \hspace{0.55cm} & \hspace{0.4cm} $\bar{p}_{N21}$ \hspace{0.4cm} & \hspace{0.4cm} $\bar{p}_{N12}$ \hspace{0.4cm} & \hspace{0.4cm} $\bar{p}_{I21}$ \hspace{0.4cm} & \hspace{0.4cm} $\bar{p}_{I12}$ \hspace{0.4cm} \\
Adiabaticity &             &                      &        &     & & \\
\hline
Adiabatic H crossing & $|U_{e3}|^2$ &  $|U_{e2}|^2$  & $|\bar{U}_{e1}|^2$  & $|\bar{U}_{e2}|^2$ & $|\bar{U}_{e3}|^2$  &  $|\bar{U}_{e3}|^2$  \\
Nonadiabatic H crossing& $|U_{e2}|^2$ &  $|U_{e2}|^2$  & $|\bar{U}_{e1}|^2$  & $|\bar{U}_{e2}|^2$  & $|\bar{U}_{e1}|^2$  &  $|\bar{U}_{e2}|^2$  \\

\hline
\end{tabular}
\label{Tabzero}
\vglue 0.5cm

\caption[aaa]{The electron neutrino and antineutrino survival factors 
$p$ and $\bar{p}$ are presented for adiabatic and nonadiabatic H level 
crossings. Note that the apparent duplication due to the 
adiabatic and non-adiabatic H level crossing in the columns of 
$p_{I}$, $\bar{p}_{N21}$, and $\bar{p}_{N12}$ are 
superficial for flavor conversion in supernova 
because of no H level crossing.
}
\label{ad-nadH}
%\vglue -0.2cm
\end{table}
%%%%%%%%%%%%%%%%% Table I %%%%%%%%%%%%%%%%%%%%

From the viewpoint of neutrino flavor transformation, however, 
there are enormous degeneracies in the 32 cases. 
First of all, the duplication due to the 
adiabatic and non-adiabatic H level crossing in the columns of 
$p_{I}$, $\bar{p}_{N21}$, and $\bar{p}_{N12}$ 
(Fig.~\ref{levelcross}b, Fig.~\ref{levelcross}c, Fig.~\ref{levelcross}e)
are superficial because of no H level crossing.
In fact, one can show from Table~\ref{ad-nadH} that there are only 6 
different patterns of the neutrino spectra;
\begin{eqnarray}
P_{31}: p&=&|U_{e3}|^2, \bar{p}=|\bar{U}_{e1}|^2 \nonumber \\
P_{23}: p&=&|U_{e2}|^2, \bar{p}=|\bar{U}_{e3}|^2 \nonumber \\
P_{21}: p &=& |U_{e2}|^2, \bar{p}=|\bar{U}_{e1}|^2 \nonumber \\
P_{32}: p&=&|U_{e3}|^2, \bar{p}=|\bar{U}_{e2}|^2 \nonumber \\
P_{33}: p&=&|U_{e3}|^2, \bar{p}=|\bar{U}_{e3}|^2  \nonumber \\
P_{22}: p &=& |U_{e2}|^2, \bar{p}=|\bar{U}_{e2}|^2
\label{pattern}
\end{eqnarray}
Each pattern contains several cases of $\nu$ and $\bar{\nu}$ 
mass hierarchies and (non-)adiabaticity of H resonance. 
Notice, however, that all the 32 cases must be treated as different 
scenarios from particle physics point of view. 
%For example, in neutrino inverted hierarchy the adiabatic and the non-adiabatic H resonance imply, respectively, a large and a small $s_{13}$, 
%
Despite degeneracies in the features of flavor conversion, 
each of the degenerate scenarios sometimes has  
different CPT transformation properties.

We use abbreviated notation to represent them.  For example, 
$(\nu: \text{N-AD}, \bar{\nu}: \text{I-NAD})$ implies that 
neutrinos have the normal mass hierarchy and the 
adiabatic H level crossing, and 
antineutrinos have the inverted mass hierarchy and the 
nonadiabatic H level crossing.
Then, the content of each flux pattern is:
\begin{eqnarray}
P_{31}: 
&&(\nu:  \text{N-AD}, \bar{\nu}: \text{N21-AD}), 
\underline {(\nu:  \text{N-AD}, \bar{\nu}: \text{N21-NAD})}, 
\underline {(\nu:  \text{N-AD}, \bar{\nu}: \text{I21-NAD})}  \nonumber \\
P_{23}: 
&&\underline {(\nu:  \text{N-NAD}, \bar{\nu}: \text{I21-AD})}, 
\underline {(\nu:  \text{N-NAD}, \bar{\nu}: \text{I12-AD})}, 
(\nu:  \text{I-AD}, \bar{\nu}: \text{I21-AD}), 
\nonumber \\
&&
\underline {(\nu:  \text{I-AD}, \bar{\nu}: \text{I12-AD})},
\underline {(\nu:  \text{I-NAD}, \bar{\nu}: \text{I21-AD})},
\underline {(\nu:  \text{I-NAD}, \bar{\nu}: \text{I12-AD})} \nonumber \\
P_{21}:
&&\underline {(\nu:  \text{N-NAD}, \bar{\nu}:  \text{N21-AD})}, 
(\nu:   \text{N-NAD}, \bar{\nu}:  \text{N21-NAD}),  
\underline {(\nu:   \text{N-NAD}, \bar{\nu}:  \text{I21-NAD})}, \nonumber \\
&&\underline {(\nu:   \text{I-AD}, \bar{\nu}:  \text{N21-AD})},
\underline {(\nu:   \text{I-AD}, \bar{\nu}:  \text{N21-NAD})},
\underline {(\nu:   \text{I-AD}, \bar{\nu}:  \text{I21-NAD})},\nonumber \\
&&\underline {(\nu:   \text{I-NAD}, \bar{\nu}:  \text{N21-AD})}, 
\underline {(\nu:   \text{I-NAD}, \bar{\nu}:  \text{N21-NAD})},
(\nu:   \text{I-NAD}, \bar{\nu}:  \text{I21-NAD}), \nonumber \\
P_{32}: 
&&\underline {(\nu:  \text{N-AD}, \bar{\nu}: \text{N12-AD})}, 
\underline {(\nu:  \text{N-AD}, \bar{\nu}: \text{N12-NAD})}, 
\underline {(\nu:  \text{N-AD}, \bar{\nu}: \text{I12-NAD})} \nonumber \\
P_{33}: 
&&\underline {(\nu:  \text{N-AD}, \bar{\nu}: \text{I21-AD})},
\underline {(\nu:  \text{N-AD}, \bar{\nu}: \text{I12-AD})}  \nonumber \\
P_{22}: 
&&\underline {(\nu:  \text{N-NAD}, \bar{\nu}:  \text{N12-AD})}, 
\underline {(\nu:   \text{N-NAD}, \bar{\nu}:  \text{N12-NAD})},  
\underline {(\nu:   \text{N-NAD}, \bar{\nu}:  \text{I12-NAD})}, \nonumber \\
&&\underline {(\nu:   \text{I-AD}, \bar{\nu}:  \text{N12-AD})},
\underline {(\nu:   \text{I-AD}, \bar{\nu}:  \text{N12-NAD})},
\underline {(\nu:   \text{I-AD}, \bar{\nu}:  \text{I12-NAD})}, \nonumber \\
&&\underline {(\nu:   \text{I-NAD}, \bar{\nu}:  \text{N12-AD})}, 
\underline {(\nu:   \text{I-NAD}, \bar{\nu}:  \text{N12-NAD})},
\underline {(\nu:   \text{I-NAD}, \bar{\nu}:  \text{I12-NAD})}, 
\label{pattern1}
\end{eqnarray}
where the one with (without) underline 
indicates the case with (without) CPT violation.
Several immediate comments are in order; 
Most notably, only the CPT violating cases are involved in 
the latter three patterns $P_{32}$, $P_{33}$ and $P_{22}$. 
Whereas the first three patterns $P_{31}$, $P_{23}$ and $P_{21}$ 
contain CPT conserving as well as violating cases;  
There are only 4 CPT conserving cases, two in $P_{21}$, and one 
in each of $P_{23}$ and in $P_{31}$, 
and the remaining 28 cases are CPT violating.

Therefore, if one is able to disentangle the latter three patterns 
observationally, the future supernova neutrino detection has a potential 
to discover CPT violation. 
We will discuss this possibility further in the subsequent sections.
In the rest of the patterns 
$P_{21}$, $P_{23}$ and $P_{31}$, with coexistence of 
CPT violating and CPT conserving cases, 
observation of supernova neutrinos by itself cannot 
signal CPT violation nor prove CPT invariance. However, 
there are possibilities that one can make stronger cases with 
the help of terrestrial experiments as we discuss in the next section.

\section{Case of reduced degeneracy with help of other types of experiments}
\label{reduced}

There are several cases in which CPT violation can be signaled 
more easily by combining SN $\nu$ and $\bar{\nu}$ observation 
with some other experiments. 
It occurs in particular in the case that the next generation 
accelerator \cite{JPARC,NOVA,SPL,beta} and/or the reactor 
experiments \cite{reactor} 
are able to measure $\theta_{13}$, or go down to the sensitivity to 
establish non-adiabatic H level crossing \cite{nufact}.
At the stage, it may be possible that some experiments can determine 
the neutrino mass hierarchy. For recent discussions, see e.g., \cite{recent}. 
We explore in this section what would be the effect of these additional 
inputs for uncovering CPT violation.
We do not discuss the cases in which CPT violation is already obvious 
by these additional informations, for example, the cases such as 
normal neutrino and inverted antineutrino mass hierarchies, or  
the measured values of $\theta_{ij}$ and $\bar{\theta}_{ij}$ 
differ with each other at a high confidence level.

\subsection{Detection of $\theta_{13}$ in reactor and accelerator experiments}
\label{adiab}

If the next generation accelerator experiments 
and/or the reactor measurement 
succeed to detect the effect of non-vanishing $\theta_{13}$ and 
$\bar{\theta}_{13}$, respectively, it means that 
$P_{H} = \bar{P}_{H}=0$. Then, the degeneracy shrinks enormously. 
In each pattern of masses and level crossings,  the cases which remain are:
\begin{eqnarray}
P_{31}: 
&&(\nu:  \text{N-AD}, \bar{\nu}: \text{N21-AD}), \nonumber\\
P_{23}: 
&&(\nu:  \text{I-AD}, \bar{\nu}: \text{I21-AD}), 
\underline {(\nu:  \text{I-AD}, \bar{\nu}: \text{I12-AD})},\nonumber\\
P_{21}:
&&\underline {(\nu:   \text{I-AD}, \bar{\nu}:  \text{N21-AD})}, \nonumber\\
P_{32}: 
&&\underline {(\nu:  \text{N-AD}, \bar{\nu}: \text{N12-AD})},  \nonumber\\
P_{33}: 
&&\underline {(\nu:  \text{N-AD}, \bar{\nu}: \text{I21-AD})},
\underline {(\nu:  \text{N-AD}, \bar{\nu}: \text{I12-AD})}, \nonumber\\
P_{22}: 
&&\underline {(\nu:   \text{I-AD}, \bar{\nu}:  \text{N12-AD})}.
\label{pattern2}
\end{eqnarray}
Novel feature of (\ref{pattern2}) is that the patterns 
$P_{21}$ and $P_{31}$ now contain only the CPT violating and 
CPT conserving cases, respectively.
In this case it is sufficient to exclude the patterns 
$P_{23}$ and $P_{31}$ to establish CPT violation.

\subsection{No signal of $\theta_{13}$ in future terrestrial experiments}
\label{nonadiab}

Suppose that, instead of positive detection which was assumed 
in the above, no indication for nonzero $\theta_{13}$ and 
$\bar{\theta}_{13}$ is obtained by the next generation accelerator 
and reactor experiments. 
If it continues to be true to the extreme sensitivity reachable by 
neutrino factory, it would imply that $P_{H} = \bar{P}_{H}=1$. 
Then, the degeneracy again decreases enormously, but in a quite 
different way of adiabatic $H$ resonances, 
\begin{eqnarray}
P_{21}:
&&(\nu:   \text{N-NAD}, \bar{\nu}:  \text{N21-NAD}),  
\underline {(\nu:   \text{N-NAD}, \bar{\nu}:  \text{I21-NAD})}, \nonumber \\
&&\underline {(\nu:   \text{I-NAD}, \bar{\nu}:  \text{N21-NAD})},
(\nu:   \text{I-NAD}, \bar{\nu}:  \text{I21-NAD}), \nonumber \\
P_{22}: 
&&\underline {(\nu:   \text{N-NAD}, \bar{\nu}:  \text{N12-NAD})},  
\underline {(\nu:   \text{N-NAD}, \bar{\nu}:  \text{I12-NAD})}, \nonumber \\
&&\underline {(\nu:   \text{I-NAD}, \bar{\nu}:  \text{N12-NAD})},
\underline {(\nu:   \text{I-NAD}, \bar{\nu}:  \text{I12-NAD})}, \nonumber \\
&&
\hskip -1.2cm
P_{23}, P_{31},  P_{32}, P_{33}: 
 \text{no case remains} 
\label{pattern3}
\end{eqnarray}
The only two patterns, $P_{21}$ and $P_{22}$ are allowed. 
Rejection of $P_{21}$ or confirmation of $P_{22}$ implies CPT violation.

\subsection{The normal $\nu$ and $\bar{\nu}$ mass hierarchies}
\label{normal}

If the neutrino and antineutrino mass hierarchies are both normal 
($\Delta m^2_{31} > 0$) and if the value of $\theta_{13}$ is not known, 
only four flux patterns remain; 
\begin{eqnarray}
P_{31}: 
&&(\nu:  \text{N-AD}, \bar{\nu}: \text{N21-AD}), 
\underline {(\nu:  \text{N-AD}, \bar{\nu}: \text{N21-NAD})}, \nonumber\\
P_{21}:
&&\underline {(\nu:  \text{N-NAD}, \bar{\nu}:  \text{N21-AD})}, 
(\nu:   \text{N-NAD}, \bar{\nu}:  \text{N21-NAD}), \nonumber\\
P_{32}: 
&&\underline {(\nu:  \text{N-AD}, \bar{\nu}: \text{N12-AD})}, 
\underline {(\nu:  \text{N-AD}, \bar{\nu}: \text{N12-NAD})}, \nonumber\\
P_{22}: 
&&\underline {(\nu:  \text{N-NAD}, \bar{\nu}:  \text{N12-AD})}, 
\underline {(\nu:   \text{N-NAD}, \bar{\nu}:  \text{N12-NAD})}.
\label{pattern4}
\end{eqnarray}
Rejection of the flux patterns $P_{31}$ and $P_{21}$, or 
confirmation of the patterns $P_{32}$ and $P_{22}$ establishes 
CPT violation.\footnote{
%%%%%%%%%%%%%% footnote %%%%%%%%%%%%%%%%%
Notice that the logic here is even if the sign of $\Delta m^2_{31}$ is measured 
only for neutrinos, for example, we assume the same sign for antineutrinos 
and yet the analysis can signal CPT violation.
}

\subsection{The inverted $\nu$ and $\bar{\nu}$ mass hierarchies}
\label{inverted}

If the neutrino and antineutrino mass hierarchies are both inverted type 
($\Delta m^2_{31} < 0$) and if the value of $\theta_{13}$ is not known, 
only three flux patterns remain.  
\begin{eqnarray}
P_{23}: 
&&(\nu:  \text{I-AD}, \bar{\nu}: \text{I21-AD}), 
\underline {(\nu:  \text{I-AD}, \bar{\nu}: \text{I12-AD})},
\nonumber \\
&&
\underline {(\nu:  \text{I-NAD}, \bar{\nu}: \text{I21-AD})},
\underline {(\nu:  \text{I-NAD}, \bar{\nu}: \text{I12-AD})} 
\nonumber\\
P_{21}:
&&
\underline {(\nu:   \text{I-AD}, \bar{\nu}:  \text{I21-NAD})},
(\nu:   \text{I-NAD}, \bar{\nu}:  \text{I21-NAD}), 
 \nonumber\\
P_{22}: 
&&
\underline {(\nu:   \text{I-AD}, \bar{\nu}:  \text{I12-NAD})},
\underline {(\nu:   \text{I-NAD}, \bar{\nu}:  \text{I12-NAD})}.
\label{pattern5}
\end{eqnarray}
To single out the pattern $P_{22}$ appears to be the easiest way to 
demonstrate CPT violation.

Under the assumptions made in Secs.~\ref{nonadiab}, \ref{normal}, 
and \ref{inverted}, CPT violation, once demonstrated, implies 
the reactor-inverted antineutrino mass hierarchy, $\bar{m}_2 <  \bar{m}_1$. 
Or, equivalently, $\bar{\theta}_{12}$ is in the dark side. 
Thus, supernova neutrinos can in principle have sensitivity to 
the $\bar{\theta}_{12}$ light-side vs. dark-side confusion.
Below, we examine the cases of additional inputs combined.

%\newpage

\subsection{Adiabatic H resonance and the normal or the inverted mass hierarchies}
\label{adiab_NI}

If the H resonance is adiabatic, and 
if the neutrino and antineutrino mass hierarchies are both normal 
only two flux patterns remain;
\begin{eqnarray}
P_{31}: 
&&(\nu:  \text{N-AD}, \bar{\nu}: \text{N21-AD}), \nonumber\\
P_{32}: 
&&\underline {(\nu:  \text{N-AD}, \bar{\nu}: \text{N12-AD})}.
\label{pattern6}
\end{eqnarray}

If the H resonance is adiabatic, and 
if the neutrino and antineutrino mass hierarchies are both inverted 
the allowed flux pattern is unique 
\begin{eqnarray}
P_{23}: 
&&(\nu:  \text{I-AD}, \bar{\nu}: \text{I21-AD}), 
\underline {(\nu:  \text{I-AD}, \bar{\nu}: \text{I12-AD})}. 
\label{pattern7}
\end{eqnarray}
In this case, there is no way of telling whether CPT is violated or not in our method.

\subsection{Non-adiabatic H resonance and the normal or the inverted mass hierarchies}
\label{nonadiab_NI}

If the H resonance is non-adiabatic, distinction between mass hierarchies, 
normal vs. inverted does not make difference in the allowed flux patterns; 
\begin{eqnarray}
P_{21}:
&&
(\nu:   \text{X-NAD}, \bar{\nu}:  \text{X21-NAD}), \nonumber\\
P_{22}: 
&&
\underline {(\nu:   \text{X-NAD}, \bar{\nu}:  \text{X12-NAD})}.
\label{pattern8}
\end{eqnarray}
where X can be N (normal) or I (inverted). 
Note that the two flux patterns are different from (\ref{pattern6}).

\section{Characteristic features of pattern dependent neutrino fluxes}
\label{characteristic}

 We now discuss the characteristic features of neutrino spectra of 
$e$, $\bar{e}$, and $x$ flavors in each classified patterns of 
neutrino flavor transformation in supernova. 
%Through such studies we may be able to obtain hints of how to distinguish them experimentally. 
%
%It may also help the readers to understand how the spectra predicted by various patterns differ qualitatively, and may give a feeling of to what extent they can be distinguished experimentally. 
%
So far we have formulated, in a generic way, the method for 
testing CPT violation with supernova neutrinos.
In the rest of this paper, we concentrate on testing CPT violation 
caused by difference between neutrino and antineutrino mass 
patterns as well as their mixing angles $\theta_{13}$ and $\bar{\theta}_{13}$. 
We take $\theta_{12} = \bar{\theta}_{12}$ in the following analysis. 
(Note that all the mixing angles are in the first octant in our convention, 
and $\theta_{23}$ does not come into play.)
In some cases based on the Garching simulation we need enormous 
accuracies of less than a few \% to distinguish between varrious spectral
patterns. (See Sec.~\ref{howwell}.) 
In such cases there is no hope of establishing CPT violation 
if the effects caused by small differences in mixing angle 
$\theta_{12}$ in $\nu$ and $\bar{\nu}$ 
sectors and the one from neutrino mass pattern coexist.

Though the current bounds on difference between 
$\theta_{12}$ and  $\bar{\theta}_{12}$ are rather mild ones 
it is quite possible that the room between them will be tighten 
up as the KamLAND experiment proceeds.  
The choice $\theta_{12} = \bar{\theta}_{12}$ 
could become mandatory if the low-energy solar 
neutrino measurement \cite{nakahata} and the dedicated 
reactor  $\bar{\theta}_{12}$ experiments 
\cite{MNTZ,goswami} are both realized.

\subsection{Characteristic features of spectral patterns of neutrino fluxes}

\begin{figure}[htbp]
\vglue 0.2cm
\begin{center}
\includegraphics[width=0.6\textwidth]{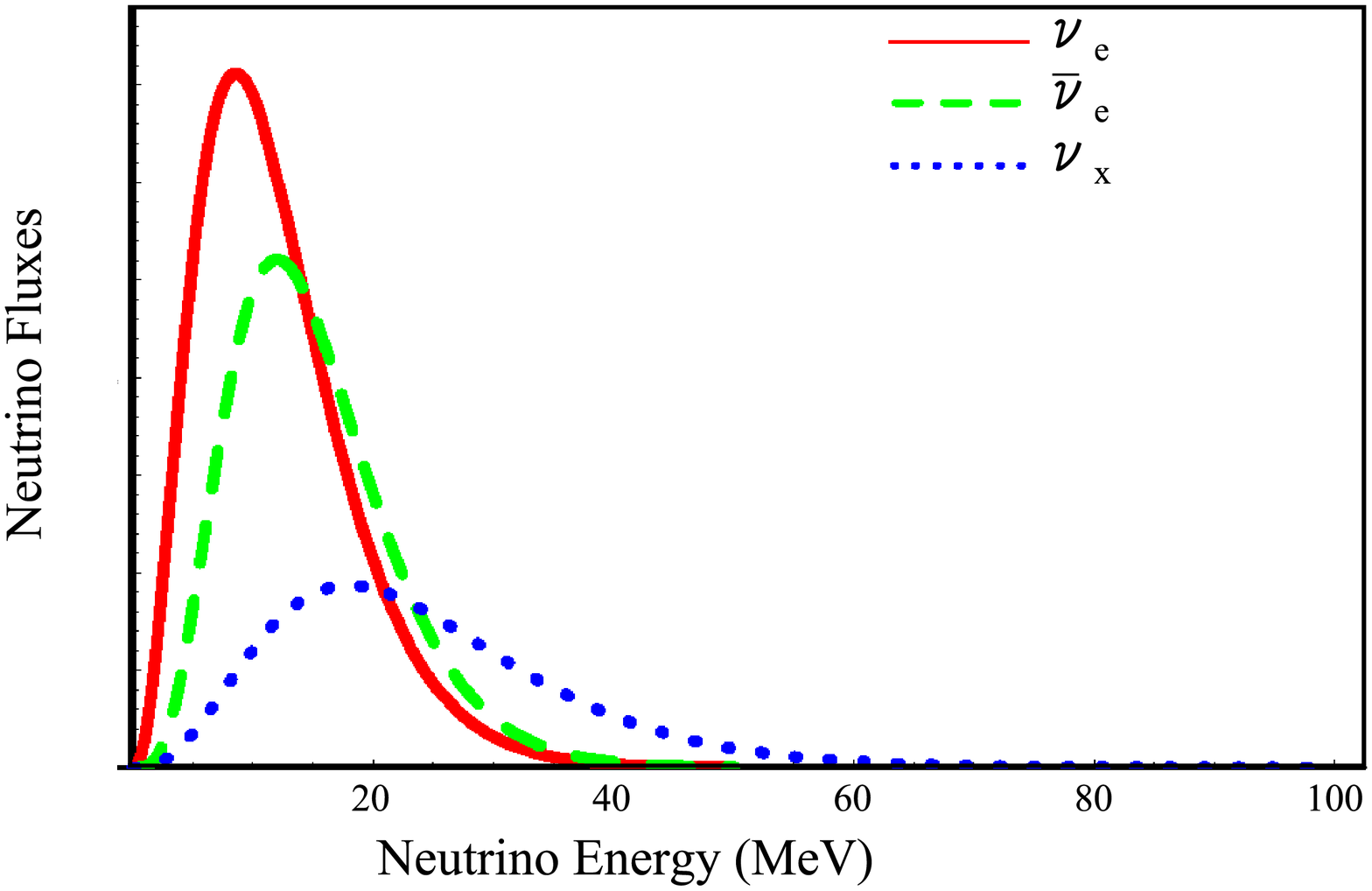}
\end{center}
%\vglue -0.2cm
\caption{The primary energy spectra of three effective neutrino species 
$\nu_{e}$ (red solid curve), 
$\bar{\nu}_{e}$ (green dashed curve),  and 
$\nu_{x}$ (blue dotted curve) of neutrinos just outside the 
neutrino sphere are shown. The flux model based on 
the Livermore simulation whose parameters are given in 
Table~\ref{parameters} in Sec.~\ref{howwell}. 
The absolute normalization is arbitrary. 
}
\label{primary}
\end{figure}

To give the readers a feeling if the neutrino spectra that arise in the six 
different patterns can be distinguished, we give an illustration using a 
model flux based on Livermore simulation \cite{livermore}. 
The parameters that characterize spectral form of the flux are given in 
Table~\ref{parameters} in Sec.~\ref{howwell}, 
where comparison between results with the other two flux models 
based on the Garching simulation is carried out. 
In this subsection we employ the Livermore flux because it is, at least, 
most suitable for illustrative purpose, having clear differences among 
spectral shapes of three effective neutrino species, as shown in 
Fig.~\ref{primary}. 
We note that the results with the Livermore parameters are very 
similar to the ones with the pinched Fermi-Dirac distribution used in 
a vast amount of literatures, e.g., 
in \cite{dighe-smi,MNTV}.

%%%%%%%%%%%% FIGURE I %%%%%%%%%%%%%%%
\vglue 0.2cm
%%%%%%%%%%%% FIGURE I %%%%%%%%%%%%%%%

%%%%%%%%%%%% FIGURE I %%%%%%%%%%%%%%%
\begin{figure}[htbp]
\vglue 0.3cm
\begin{center}
\includegraphics[width=0.98\textwidth]{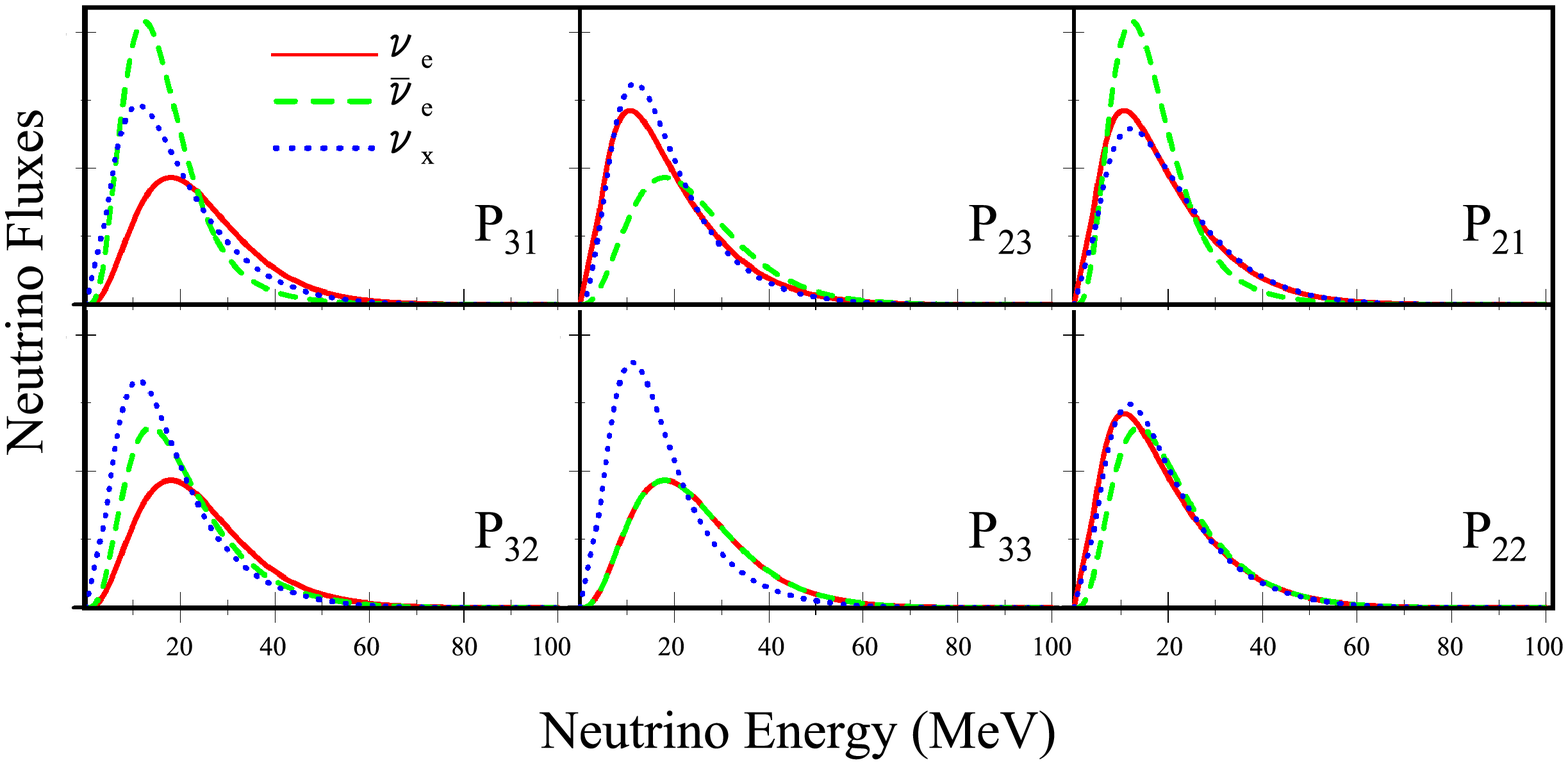}
\end{center}
\vglue -0.2cm
\caption{The energy spectra of the three effective neutrino species 
$\nu_{e}$ (red solid curve), 
$\bar{\nu}_{e}$ (green dashed curve),  and 
$\nu_{x}$ (blue dotted curve) at terrestrial detectors 
corresponding to six different patterns of neutrino flavor conversion 
$P_{31}$ to $P_{22}$ defined in (\ref{pattern}) are presented. 
The Livermore flux model with the same values of parameters 
as in Fig.~\ref{primary} is used.  
Although an over-all absolute normalization is arbitrary, the relative 
normalization between the six patterns are meaningful. 
}
\label{spectra_L}
\end{figure}
%%%%%%%%%%%% FIGURE I %%%%%%%%%%%%%%%

We draw in Fig.~\ref{spectra_L} 
the spectra of the three effective neutrino species 
$\nu_{e}$, $\bar{\nu}_{e}$, and $\nu_{x}$ of six different patterns of 
$P_{31}$ to $P_{22}$ defined in (\ref{pattern}). 
One can recognize that the six patterns of neutrino flavor conversion 
in supernova are quite different with each other partly due to the 
``optimistic'' choice of the parameters. 
The difference in spectral patterns in three species of neutrinos 
is the key to discriminate six different scenarios of flavor transformation. 
For more quantitative understanding, we give in 
Table~\ref{Livermore} the average values of energies 
$\langle E_{\alpha} \rangle$ ($\alpha =$ e, $\bar{e}$ and x), 
and the width parameter 
$\langle \Delta E_{\alpha} \rangle \equiv 
\sqrt{\langle E_{\alpha}^2 \rangle - \langle E_{\alpha} \rangle^2}$ 
which is also used in \cite{luna-smi2}. 
To make the distinction among the six patterns clearer 
we also give in Table~\ref{Livermore} the ratios 
$\langle E_{\beta} \rangle / \langle E_{\bar{e}} \rangle$ and 
$\langle \Delta E_{\beta} \rangle / \langle \Delta E_{\bar{e}} \rangle$ 
for $\beta =$ e and x, assuming that the denominators would be the 
best determined parameters. 
The distinction among the different flavor conversion pattern is obvious.

%%%%%%%%%%%%%%%%% Table II Livermore %%%%%%%%%%%%%%%%%%%%
\begin{table}
\vglue 0.5cm
\begin{tabular}{|c|c|c|c|c|c|c|c|c|c|c|}
\hline
Spectral  & \hspace{0.0cm} 
$\langle E_{e} \rangle$ \hspace{0.0cm} & \hspace{0.0cm} 
$\langle E_{\bar{e}} \rangle$ \hspace{0.0cm} & \hspace{0.0cm} 
$\langle E_{x} \rangle$ \hspace{0.0cm} & \hspace{-0.1cm} 
$\langle \Delta E_{e} \rangle$ \hspace{-0.1cm} & \hspace{-0.1cm} 
$\langle \Delta E_{\bar{e}} \rangle$ \hspace{-0.1cm} & \hspace{-0.1cm} 
$\langle \Delta E_{x} \rangle$ \hspace{-0.1cm} & \hspace{0.1cm} 
$\frac{\langle E_{e} \rangle}{\langle E_{\bar{e}} \rangle}$ \hspace{0.1cm} & \hspace{0.1cm} 
$\frac{\langle E_{x} \rangle}{\langle E_{\bar{e}} \rangle}$ \hspace{0.1cm} & \hspace{0.0cm} 
$\frac{\langle \Delta E_{e} \rangle}{\langle \Delta E_{\bar{e}} \rangle}$ \hspace{0.0cm} & \hspace{0.0cm} 
$\frac{\langle \Delta E_{x} \rangle}{\langle \Delta E_{\bar{e}} \rangle}$ \hspace{0.0cm} 
 \\
patterns &             &            &  &          &        &    & &  & & \\
\hline
$P_{31}$ & 24.0 &  16.8  & 18.6 & 12.0 &  8.8  &  11.3  &  1.43 &  1.10 & 1.36 & 1.29 \\
$P_{23}$ & 18.7 &  24.0  & 18.1 & 11.6 & 12.0  & 10.7 & 0.78 &  0.75 &  0.96 &  0.89 \\
$P_{21}$ & 18.7 &  16.8  & 19.6  & 11.6 & 8.8 &  11.6  & 1.11 & 1.17 & 1.31 & 1.32 \\
\hline
$P_{32}$ & 24.0 & 20.5 & 17.7  & 12.0 & 11.1 &  10.7  & 1.17 & 0.86 & 1.07 & 0.96 \\
$P_{33}$ & 24.0 & 24.0 & 17.1 & 12.0 & 12.0 & 10.3 & 1.0 & 0.71 &  1.0 &  0.86 \\
$P_{22}$ & 18.7 &  20.5 & 17.7 & 11.6 & 11.2 & 12.6 & 0.91 & 0.86 & 1.04 & 1.13 \\
\hline
\end{tabular}
\vglue 0.5cm

\caption[aaa]{
The averaged energies and the width parameters of three species of 
neutrinos are presented by using the spectra based on the 
Livermore simulation with the parameters given in 
Table~\ref{parameters}. 
The upper three patterns, $P_{31}$, $P_{23}$, and $P_{21}$ 
contain CPT conserving as well as violating cases, whereas the lower three patterns 
$P_{32}$, $P_{33}$, and $P_{22}$ consist solely of CPT violating ones.
}
\label{Livermore}
\end{table}
\vglue 0.2cm
%%%%%%%%%%%%%%% Table II Livermore %%%%%%%%%%%%%%%%%%

Notice that the upper three patterns, $P_{31}$, $P_{23}$, and $P_{21}$ 
contain CPT conserving as well as violating cases, whereas the lower 
three patterns 
$P_{32}$, $P_{33}$, and $P_{22}$ consist solely of CPT violating ones.
Therefore, the above discussion applies, upon restriction to the upper 
three cases, to the conventional CPT conserving cases as well.

We make comments on some notable features of the results 
presented in Table~\ref{Livermore}. 

\begin{itemize}

\item
CPT conserving cases

It may be instructive to understand the feature of 
the CPT conserving cases contained in the patterns 
$P_{31}$, $P_{23}$, and $P_{21}$.  
In the pattern $P_{31}$ 
(the case of normal hierarchy and adiabatic H resonance) 
the H resonance is in the neutrino channel, and 
$\langle E_{e} \rangle / \langle E_{\bar{e}} \rangle > 1$ because 
$\nu_e$ at the terrestrial detector is dominantly composed by 
$\nu_x$ at the neutrinosphere \cite{MN90}.
In the pattern $P_{23}$ 
(the case of inverted hierarchy and adiabatic H resonance) 
the H resonance is in the antineutrino channel, and therefore 
$\langle E_{e} \rangle / \langle E_{\bar{e}} \rangle < 1$. 
The distinction between $P_{31}$ and $P_{23}$ allows 
to distinguish the normal and the inverted mass hierarchies 
\cite{MN_inverted}. 
In the pattern $P_{21}$ 
(the case of non-adiabatic H resonance) 
$\nu_e$ ($\bar{\nu}_e$) at the earth are superposition of 
$\nu_e$ ($\bar{\nu}_e$) and $\nu_x$ at the neutrinosphere. 
These features are extensively discussed by many authors 
\cite{MN_inverted,dighe-smi,dutta,todai,MNTV,barger,luna-smi1,luna-smi2}.

\item
CPT violating cases

As one can recognize from Table~\ref{Livermore} there is a general 
tendency that 
$\langle E_{x} \rangle / \langle E_{\bar{e}} \rangle$ is small in 
CPT violating cases. 
Unfortunately, it does not guarantee unique identification of them 
because of the similar small ratio of $P_{23}$.
But the latter has distinctive feature that all the ratios 
$\langle E_{\alpha} \rangle / \langle E_{\bar{e}} \rangle$ and 
$\langle \Delta E_{\alpha} \rangle / \langle \Delta E_{\bar{e}} \rangle$ 
($\alpha = e, x$) 
are smaller than unity. The feature may allow unique identification 
of $P_{23}$, and hence that of CPT violating cases by elimination.

\end{itemize}

\subsection{Approximate Analytic Expression of the Flux Composition}
\label{analytic}

To facilitate clear understanding and to complement the 
drawing of figures we give approximate expressions of fluxes. 
They will help understanding the features of Fig.~\ref{spectra_L}. 
We use the approximations $s_{13} \ll 1$ and $\bar{s}_{13} \ll 1$ 
(assuming that the former is true) which are numerically valid.  
Notice that it {\it does not} mean that we restrict to the case of 
non-adiabatic H-level crossing. 
The approximation applies to the expressions of fluxes 
after taking $P_{H} =0$ or $P_{H} =1$ etc. to merely simplify the expressions.

\begin{itemize}
\item 
{\bf Pattern $P_{31}$}
\end{itemize}

In this case $p = |U_{e3}|^2 = s^2_{13}$ and 
$\bar{p} = |\bar{U}_{e1}|^2 = \bar{c}^2_{12} \bar{c}^2_{13}$.
Then, the $\nu_e$, $\bar{\nu}_{e}$, and $\nu_x$ spectra are given by:
\begin{eqnarray}
F_e &=& 
s^2_{13} F_e^0 + c^2_{13} F_x^0 
\approx F_x^0,
\nonumber \\
F_{\bar{e}} &=&
\bar{c}^2_{12} \bar{c}^2_{13} F_{\bar{e}}^0 + 
(1- \bar{c}^2_{12} \bar{c}^2_{13}) F_x^0 
\approx \bar{c}^2_{12} F_{\bar{e}}^0 + \bar{s}^2_{12} F_x^0, 
\nonumber \\
4 F_x &=& 
c^2_{13}F_e^0 + (1- \bar{c}^2_{12} \bar{c}^2_{13}) F_{\bar{e}}^0 + 
(2 + s^2_{13} + \bar{c}^2_{12} \bar{c}^2_{13}) F_x^0 
\nonumber \\
&\approx& F_e^0 + \bar{s}^2_{12} F_{\bar{e}}^0 + (2 + \bar{c}^2_{12}) F_x^0.
\label{F_31}
\end{eqnarray}

\begin{itemize}
\item 
{\bf Pattern $P_{23}$}
\end{itemize}

In this case $p = |U_{e2}|^2 = s^2_{12} c^2_{13}$ and 
$\bar{p} = |\bar{U}_{e3}|^2 = \bar{s}^2_{13}$.
Then, the $\nu_e$, $\bar{\nu}_{e}$, and $\nu_x$ spectra are given by:

\begin{eqnarray}
F_{e} &=& 
s^2_{12} c^2_{13} F_e^0 + (1- s^2_{12} c^2_{13}) F_x^0 
\approx s^2_{12} F_e^0 + c^2_{12} F_x^0, 
\nonumber \\
F_{\bar{e}}  &=& 
 \bar{s}^2_{13} F_e^0 +  \bar{c}^2_{13} F_x^0 
\approx F_x^0,
\nonumber \\
4 F_x &=& 
(1- s^2_{12} c^2_{13}) F_e^0 + \bar{c}^2_{13} F_{\bar{e}}^0 + 
(2 + s^2_{12} c^2_{13} + \bar{s}^2_{13}) F_x^0 
\nonumber \\
&\approx& c^2_{12} F_e^0 + F_{\bar{e}}^0 + (2 + s^2_{12}) F_x^0.
\label{F_23}
\end{eqnarray}

\begin{itemize}
\item 
{\bf Pattern $P_{21}$}
\end{itemize}

In this case $p = |U_{e2}|^2 = s^2_{12} c^2_{13}$ and 
$\bar{p} = |\bar{U}_{e1}|^2 = \bar{c}^2_{12} \bar{c}^2_{13}$.
Then, the $\nu_e$, $\bar{\nu}_{e}$, and $\nu_x$ spectra are given by:
\begin{eqnarray}
F_e &=& 
s^2_{12} c^2_{13} F_e^0 + (1- s^2_{12} c^2_{13}) F_x^0 
\approx s^2_{12} F_e^0 + c^2_{12} F_x^0, 
\nonumber \\
F_{\bar{e}} &=&
\bar{c}^2_{12} \bar{c}^2_{13} F_{\bar{e}}^0 + 
(1- \bar{c}^2_{12} \bar{c}^2_{13}) F_x^0 
\approx \bar{c}^2_{12} F_{\bar{e}}^0 + \bar{s}^2_{12} F_x^0,
\nonumber \\
4 F_x &=& 
(1- s^2_{12} c^2_{13}) F_e^0 + (1- \bar{c}^2_{12} \bar{c}^2_{13}) F_{\bar{e}}^0 + 
(2 + s^2_{12} c^2_{13} + \bar{c}^2_{12} \bar{c}^2_{13}) F_x^0 
\nonumber \\
&\approx& c^2_{12} F_e^0 + \bar{s}^2_{12} F_{\bar{e}}^0 + 
(2 + s^2_{12} + \bar{c}^2_{12}) F_x^0.
\label{F_21}
\end{eqnarray}

\begin{itemize}
\item 
{\bf Pattern $P_{32}$}
\end{itemize}

This case contains only CPT violating patterns. 
In this case $p = |U_{e3}|^2 = s^2_{13}$ and 
$\bar{p} = |\bar{U}_{e2}|^2 = \bar{s}^2_{12} \bar{c}^2_{13}$.
Then, the $\nu_e$, $\bar{\nu}_{e}$, and $\nu_x$ spectra are given by:
\begin{eqnarray}
F_e &=& 
s^2_{13} F_e^0 + c^2_{13} F_x^0 
\approx F_x^0,
\nonumber \\
F_{\bar{e}} &=&
\bar{s}^2_{12} \bar{c}^2_{13} F_{\bar{e}}^0 + 
(1- \bar{s}^2_{12} \bar{c}^2_{13}) F_x^0 
\approx \bar{s}^2_{12} F_{\bar{e}}^0 + \bar{c}^2_{12} F_x^0, 
\nonumber \\
4 F_x &=& 
c^2_{13}F_e^0 + (1- \bar{s}^2_{12} \bar{c}^2_{13}) F_{\bar{e}}^0 + 
(2 + s^2_{13} + \bar{s}^2_{12} \bar{c}^2_{13}) F_x^0 
\nonumber \\
&\approx& F_e^0 + \bar{c}^2_{12} F_{\bar{e}}^0 + (2 + \bar{s}^2_{12}) F_x^0.
\label{F_32}
\end{eqnarray}

\begin{itemize}
\item 
{\bf Pattern $P_{33}$}
\end{itemize}

This case contains only CPT violating patterns. 
In this case $p = |U_{e3}|^2 = s^2_{13}$ and 
$\bar{p} = |\bar{U}_{e3}|^2 = \bar{s}^2_{13}$.
Then, the $\nu_e$, $\bar{\nu}_{e}$, and $\nu_x$ spectra are given by:

\begin{eqnarray}
F_e &=& 
s^2_{13} F_e^0 + c^2_{13} F_x^0 
\approx F_x^0,
\nonumber \\
F_{\bar{e}}  &=& 
 \bar{s}^2_{13} F_e^0 +  \bar{c}^2_{13} F_x^0 
\approx F_x^0,
\nonumber \\
4 F_x &=& 
c^2_{13}F_e^0 +  \bar{c}^2_{13} F_{\bar{e}}^0 + 
(2 + s^2_{13} +  \bar{s}^2_{13}) F_x^0 
\nonumber \\
&\approx& F_e^0 +  F_{\bar{e}}^0 + 2 F_x^0.
\label{F_33}
\end{eqnarray}

\begin{itemize}
\item 
{\bf Pattern $P_{22}$}
\end{itemize}

This case contains only CPT violating patterns. 
In this case $p = |U_{e2}|^2 = s^2_{12} c^2_{13}$ and 
$\bar{p} = |\bar{U}_{e2}|^2 = \bar{s}^2_{12} \bar{c}^2_{13}$.
Then, the $\nu_e$, $\bar{\nu}_{e}$, and $\nu_x$ spectra are given by:
\begin{eqnarray}
F_e &=& 
s^2_{12} c^2_{13} F_e^0 + (1- s^2_{12} c^2_{13}) F_x^0 
\approx s^2_{12} F_e^0 + c^2_{12} F_x^0, 
\nonumber \\
F_{\bar{e}} &=&
\bar{s}^2_{12} \bar{c}^2_{13} F_{\bar{e}}^0 + 
(1- \bar{s}^2_{12} \bar{c}^2_{13}) F_x^0 
\approx \bar{s}^2_{12} F_{\bar{e}}^0 + \bar{c}^2_{12} F_x^0,
\nonumber \\
4 F_x &=& 
(1- s^2_{12} c^2_{13}) F_e^0 + (1- \bar{s}^2_{12} \bar{c}^2_{13}) F_{\bar{e}}^0 + 
(2 + s^2_{12} c^2_{13} + \bar{s}^2_{12} \bar{c}^2_{13}) F_x^0 
\nonumber \\
&\approx& c^2_{12} F_e^0 + \bar{c}^2_{12} F_{\bar{e}}^0 + 
(2 + s^2_{12} + \bar{s}^2_{12}) F_x^0.
\label{F_22}
\end{eqnarray}

%\newpage

\section{To What Extent Can Neutrino Flux Patterns Be Discriminated?}
\label{howwell}

In this section, we briefly discuss to what extent the flux patterns 
predicted by six cases from $P_{31}$ to $P_{22}$ can be discriminated 
observationally.
Our discussion cannot be a quantitative one because of the lack of 
``standard supernova model'' which has comparable accuracies 
possessed by the standard solar model.
But, it may give us a feeling of how accurate should be 
the flavor-dependent reconstruction of neutrino spectra to 
discriminate the six different patterns of flavor conversion.

We also address in Sec.\ref{help} 
the question to what extent limiting the number of possible 
flux patterns by additional inputs help identifying CPT violation.

\subsection{Model dependence of the prediction to flux spectral patterns}

To reflect the best knowledges of supernova simulation currently at hand, 
we employ in this section the parametrization of fluxes which is used by 
the Garching group to fit their data \cite{MKeil,Keil02}:
\begin{eqnarray}
F^0_{\alpha}(E) =
 \frac{\Phi_{\alpha}}{\langle E_{\alpha} \rangle}\,
 \frac{\beta_{\alpha}^{\beta_{\alpha}}}{\Gamma(\beta_{\alpha})}  
 \left(\frac{E}{\langle E_{\alpha} \rangle}\right)^{\beta_{\alpha}-1} 
 \exp\left(-\beta_{\alpha}\frac{E}{\langle E_{\alpha} \rangle}\right) \,,
\label{flux-form}
\end{eqnarray}
where $\langle E_{\alpha} \rangle$ denotes their average energy,
$\beta_{\alpha}$ is a dimensionless parameter which is related to 
the width of the spectrum and typically takes on values 3.5--6. 
For definiteness, we assume $\beta_{\nu_e}=3.5$, 
$\beta_{\nu_x}=4$, and $\beta_{\bar{\nu}_{e}}=5$.
For the sake of comparison and to reveal dependence on 
supernova simulations we examine the three different sets of parameters, 
the same ones used in \cite{tomas_etal}. 
They are the fluxes obtained from the Livermore simulation~\cite{livermore}, 
and typical two model fluxes based on simulation done by the Garching
group~\cite{garching}. The three model flux parameters are given in 
Table~\ref{parameters} as L, G1 and G2.

\vglue 0.1cm

\begin{table}[ht]
\begin{tabular}{cccccc}
\hline
Model & $\langle E_0(\nu_e) \rangle$ & $\langle E_0(\bar{\nu}_{e}) \rangle$ &
$\langle E_0(\nu_x) \rangle$ & {\large $\frac{\Phi_0(\nu_e)}{\Phi_0(\nu_x)}$} &
{\large $\frac{\Phi_0(\bar{\nu}_{e})}{\Phi_0(\nu_x)}$}\\
\hline
L & 12 & 15 & 24 & 2.0 & 1.6 \\
G1 & 12 & 15 & 18 & 0.8 & 0.8 \\
G2 & 12 & 15 & 15 & 0.5 & 0.5 \\
\hline
\end{tabular}
\vglue 0.3cm
\caption{The parameters of the primary neutrino spectra models
  motivated from SN simulations of the Garching (G1, G2) and the
  Livermore (L) groups, the same as used in \cite{tomas_etal}. 
  We assume $\beta_{\nu_e}=3.5$, $\beta_{\nu_x}=4$, and $\beta_{\bar{\nu}_{e}}=5$.
  } 
\label{parameters}
\end{table} 

% \vglue -0.3cm

\begin{figure}[htbp]
\vglue 0.2cm
\begin{center}
\hglue 1.0cm
\includegraphics[width=0.98\textwidth]{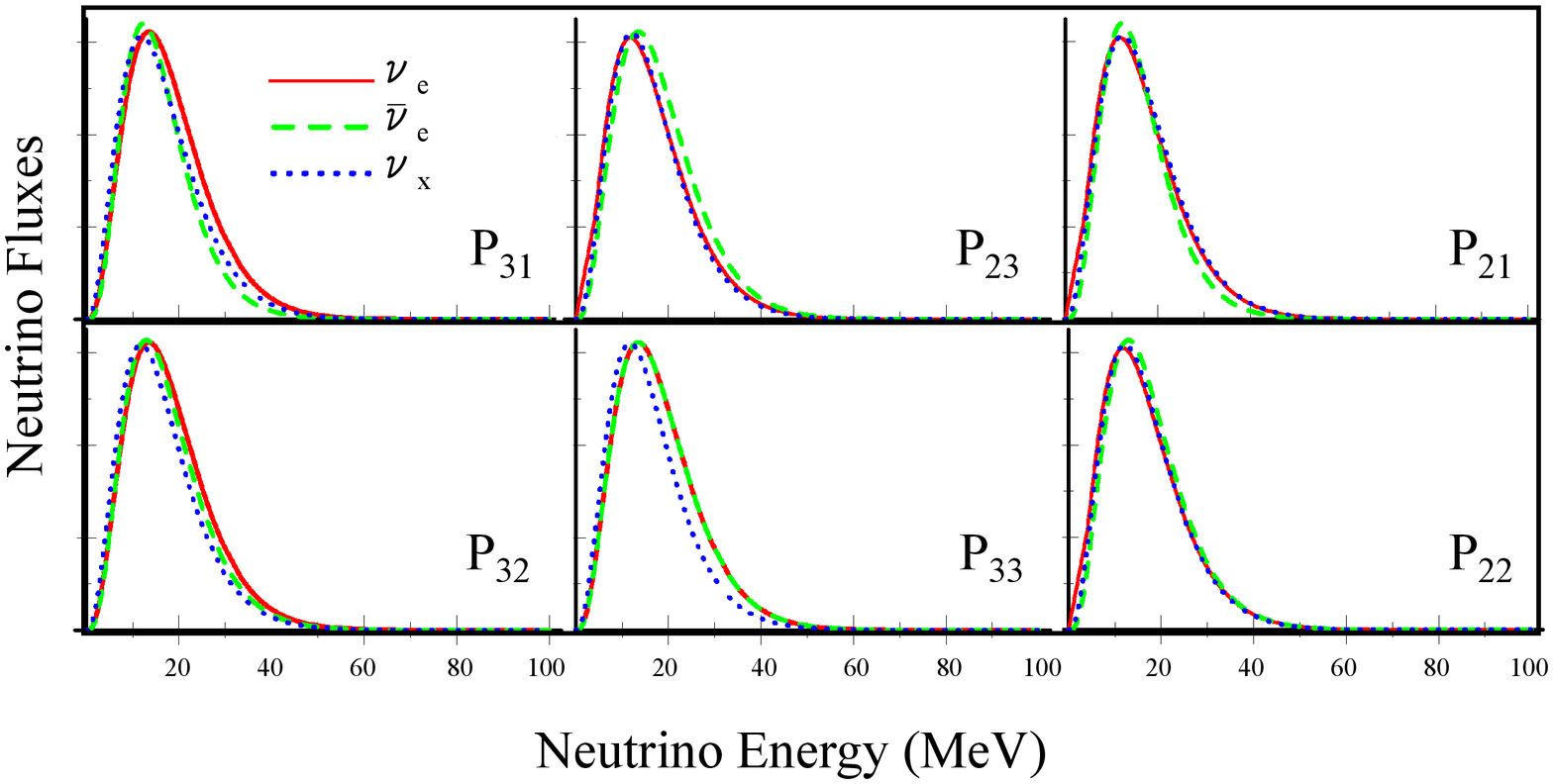}
\vglue -0.6cm
\end{center}
\begin{center}
\vglue -0.4cm
\hglue 2.0cm
\includegraphics[width=0.98\textwidth]{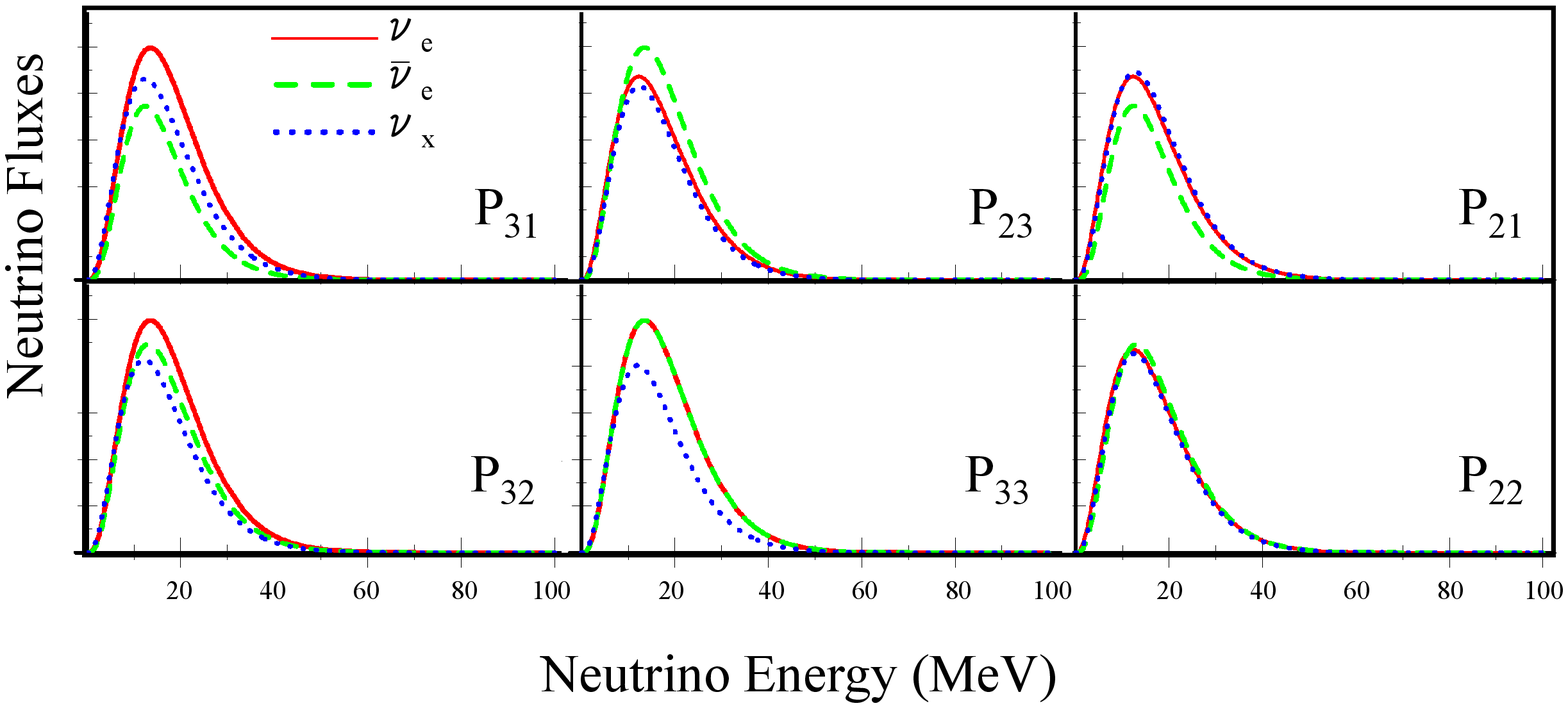}
\end{center}
\vglue -1.0cm
\caption{The energy spectra of the three effective neutrino species 
$\nu_{e}$ (red solid curve), 
$\bar{\nu}_{e}$ (green dashed curve),  and 
$\nu_{x}$ (blue dotted curve) at terrestrial detectors 
corresponding to six different patterns of neutrino flavor conversion 
$P_{31}$ to $P_{22}$ defined in (\ref{pattern}) are presented. 
The Garching flux model parameters G1 (upper figures) 
and G2 (lower figures) are used. 
}
\label{spectra_G}
\end{figure}

In Fig.~\ref{spectra_G}, 
the spectra of the three effective neutrino species 
$\nu_{e}$, $\bar{\nu}_{e}$, and $\nu_{x}$ of six different patterns of 
$P_{31}$ to $P_{22}$ defined in (\ref{pattern}) are plotted 
by taking model parameters G1 (upper figures) and 
G2 (lower figures) in Table~\ref{parameters}. 
As one can recognize the six flavor conversion patterns is much harder 
to distinguish than the case of Livermore flux.

%%%%%%%%%%%% FIGURE I %%%%%%%%%%%%%%%
%%%%%%%%%%%% FIGURE I %%%%%%%%%%%%%%%

To obtain a hint on how much accuracy is needed to disentangle 
these six patterns we give in  Table~\ref {Garching} 
the averaged energies and width parameters of three species of neutrinos 
as done in Table~\ref {Livermore}. 
With the Garching parameters,  
typically a few \% accuracies for the determination of ratios 
$\langle E_{\alpha} \rangle / \langle E_{\bar{e}} \rangle$ and 
$\langle \Delta E_{\alpha} \rangle / \langle \Delta E_{\bar{e}} \rangle$ 
($\alpha =$ e and x) are required. 
Whereas in the case of the Livermore parameters, 
we may be able to do the job with the accuracies of $\sim$10 \%.

In addition to these quantities we have added in  Table~\ref {Garching} 
columns for the ratio of luminosity of $\nu_e$ and $\nu_x$ to $\bar{\nu}_e$. 
(We note that even if the columns are added in Table~\ref {Livermore} 
it is not informative because of the equi-partition of luminosity 
between three species.) 
It should not be too difficult to distinguish among the six patterns 
if the luminosity are measured at the level of a few \% level (G1) 
and 5-10\% level (G2). 
%In particular the distinction is easier in the case of model flux G2. 

%%%%%%%%%%%%%%%%% Table II G1 & G2 %%%%%%%%%%%%%%%%%%%%
\begin{table}
\vglue 0.5cm
\begin{tabular}{|c|c|c|c|c|c|c|c|c|c|c|c|c|}
\hline
Spectral  & \hspace{0.0cm} 
$\langle E_{e} \rangle$ \hspace{0.0cm} & \hspace{0.0cm} 
$\langle E_{\bar{e}} \rangle$ \hspace{0.0cm} & \hspace{0.0cm} 
$\langle E_{x} \rangle$ \hspace{0.0cm} & \hspace{-0.1cm} 
$\langle \Delta E_{e} \rangle$ \hspace{-0.1cm} & \hspace{-0.1cm} 
$\langle \Delta E_{\bar{e}} \rangle$ \hspace{-0.1cm} & \hspace{-0.1cm} 
$\langle \Delta E_{x} \rangle$ \hspace{-0.1cm} & \hspace{0.1cm} 
$\frac{\langle E_{e} \rangle}{\langle E_{\bar{e}} \rangle}$ \hspace{0.1cm} & \hspace{0.1cm} 
$\frac{\langle E_{x} \rangle}{\langle E_{\bar{e}} \rangle}$ \hspace{0.1cm} & \hspace{0.0cm} 
$\frac{\langle \Delta E_{e} \rangle}{\langle \Delta E_{\bar{e}} \rangle}$ \hspace{0.0cm} & \hspace{0.0cm} 
$\frac{\langle \Delta E_{x} \rangle}{\langle \Delta E_{\bar{e}} \rangle}$ \hspace{0.0cm} & \hspace{0.1cm}
$\frac{\langle L_{e} \rangle}{\langle L_{\bar{e}} \rangle}$ \hspace{0.1cm} &\hspace{0.1cm} 
$\frac{\langle L_{x} \rangle}{\langle L_{\bar{e}} \rangle}$ \hspace{0.1cm} \\
patterns &             &            &  &          &        &    & &  & & & & \\
\hline
$P_{31}$ & 18.0 &  16.0  & 16.5 & 9.0 & 7.7 & 8.7 &  1.12 &  1.03 & 1.17 & 1.13 & 1.31 & 1.13 \\
$P_{23}$ & 16.5 &  18.0  & 16.4 & 8.8 & 9.0 & 8.5 &  0.92&  0.91 &  0.98 &  0.94 & 0.87 & 0.83 \\
$P_{21}$ & 16.5 &  16.0  & 16.9 & 8.8 & 7.7 & 8.8 &  1.03 & 1.06 & 1.15 & 1.15 & 1.14 & 1.17 \\
\hline
$P_{32}$ & 18.0 &  17.3  & 16.2 & 9.0 & 8.6 & 8.5 &  1.04 & 0.94 & 1.05 & 0.99 & 1.11 & 0.91 \\
$P_{33}$ & 18.0 &  18.0  & 16.0 & 9.0 & 8.4 & 8.4 &  1.0 & 0.89 &  1.0 &  0.93 & 1.0 & 0.80 \\
$P_{22}$ & 16.5 &  17.3  & 16.9 & 8.8 & 8.6 & 8.1 &  0.96 & 0.98 & 1.03 & 0.94 & 0.96 & 0.95 \\
\hline
\end{tabular}
\vglue 0.8cm
\begin{tabular}{|c|c|c|c|c|c|c|c|c|c|c|c|c|}
\hline
Spectral  & \hspace{0.0cm} 
$\langle E_{e} \rangle$ \hspace{0.0cm} & \hspace{0.0cm} 
$\langle E_{\bar{e}} \rangle$ \hspace{0.0cm} & \hspace{0.0cm} 
$\langle E_{x} \rangle$ \hspace{0.0cm} & \hspace{-0.1cm} 
$\langle \Delta E_{e} \rangle$ \hspace{-0.1cm} & \hspace{-0.1cm} 
$\langle \Delta E_{\bar{e}} \rangle$ \hspace{-0.1cm} & \hspace{-0.1cm} 
$\langle \Delta E_{x} \rangle$ \hspace{-0.1cm} & \hspace{0.1cm} 
$\frac{\langle E_{e} \rangle}{\langle E_{\bar{e}} \rangle}$ \hspace{0.1cm} & \hspace{0.1cm} 
$\frac{\langle E_{x} \rangle}{\langle E_{\bar{e}} \rangle}$ \hspace{0.1cm} & \hspace{0.0cm} 
$\frac{\langle \Delta E_{e} \rangle}{\langle \Delta E_{\bar{e}} \rangle}$ \hspace{0.0cm} & \hspace{0.0cm} 
$\frac{\langle \Delta E_{x} \rangle}{\langle \Delta E_{\bar{e}} \rangle}$ \hspace{0.0cm} & \hspace{0.1cm}
$\frac{\langle L_{e} \rangle}{\langle L_{\bar{e}} \rangle}$ \hspace{0.1cm} &\hspace{0.1cm} 
$\frac{\langle L_{x} \rangle}{\langle L_{\bar{e}} \rangle}$ \hspace{0.1cm} \\
patterns &             &            &  &          &        &    & &  & & & & \\
\hline
$P_{31}$ & 15.0 &  15.0  & 14.6 & 7.5 &  7.1  &  7.4  &  1.0 &  0.97 & 1.06 & 1.05 & 1.56 & 1.27 \\
$P_{23}$ & 14.5 &  15.0  & 14.7 & 7.4 & 7.5  & 7.3 & 0.97 &  0.98 &  0.99 &  0.98 & 0.83 & 0.77 \\
$P_{21}$ & 14.5 &  15.0  & 14.7  & 7.4 & 7.1 &  7.4  & 0.97 & 0.98 & 1.05 & 1.05 & 1.29 & 1.33 \\
\hline
$P_{32}$ & 15.0 & 15.0 & 14.5  & 7.5 & 7.4 &  7.3  & 1.0 & 0.97 & 1.02 & 0.99 & 1.17 & 0.89 \\
$P_{33}$ & 15.0 & 15.0 & 14.5 & 7.5 & 7.5 & 7.3 & 1.0 & 0.97 &  1.0 &  0.97 & 1.0 & 0.73 \\
$P_{22}$ & 14.5 &  15.0 & 15.3 & 7.4 & 7.4 & 5.9 & 0.97 & 1.02 & 1.01 & 0.79 & 0.97 & 0.94 \\
\hline
\end{tabular}
\vglue 0.5cm

\caption[aaa]{
The same as in Table~\ref{Livermore} but with the flux models 
G1 (upper table) and G2 (lower table) based on the Garching simulation 
with the parameters given in Table~\ref{parameters}. 
Two extra columns for luminosity ratios are added.
}\label{Garching}
%\vglue -0.2cm
\end{table}
%%%%%%%%%%%%%%%%% Table II G1&G2 %%%%%%%%%%%%%%%%%%%%

\subsection{How accurately can the supernova neutrino fluxes be determined?}

To give the readers some feeling we give a brief remark on 
how accurately determination of the supernova neutrino fluxes 
can be done in future observation of galactic supernovae.
We must emphasize that the method for such flux reconstruction, 
which is of great importance solely from the supernova core 
diagnostics independent of testing CPT symmetry, is not yet developed 
to a sufficient level. It is the important problem that deserves 
thorough study to identify a minimal set of detectors which are 
capable of reconstructing fluxes which can watch supernova for 
a long run of at least $\sim$50 years.

Lacking such studies we restrict ourselves to accuracy expected 
for parameters determinable with $\bar{\nu}_{e}$ observation 
in water Cherenkov detectors. 
In \cite{MNTV} it was found that the parameters of the primary flux 
$\langle E_{\bar{e}} \rangle$ and 
$\tau_{E} \equiv \langle E_{x} \rangle / \langle E_{\bar{e}} \rangle$ 
can be determined to the accuracies 1\% (4\%) and 1.5\% (9\%) 
at 3 $\sigma$ CL with Hyper-Kamiokande \cite{nakamura} 
(Super-Kamiokande), respectively. 
The accuracies found in  \cite{MNTV} correspond to the ones of 
primary fluxes but we here assume that the similar accuracies can 
be expected for the terrestrial fluxes. 
If these accuracies can be extended to $\nu_e$ flux 
(which is, however, highly nontrivial) 
it may be possible to disentangle the flux patterns expected 
in the six different patterns of flavor conversion. 
We stress that though the accuracy required for flux determination 
for identifying CPT violation is quite a demanding one, 
they are at the same level as that required 
to diagnose interior of supernova core by means of neutrinos.

\subsection{How additional inputs from accelerator and reactor experiments help identifying CPT violation?}
\label{help}

In Sec.~\ref{reduced} we have discussed which flux patterns will 
remain if the next generation accelerator and reactor experiments 
measure $\theta_{13}$ and $\bar{\theta}_{13}$ and/or 
determine the $\nu$ and $\bar{\nu}$ mass hierarchies.
We discuss in this subsection how the reduction of numbers of 
possible flux patterns help to discriminate between CPT invariant 
and non-invariant cases.

If the H level crossing is non-adiabatic, the two flux patterns 
$P_{21}$ and $P_{22}$ remain. In fact, it is the general feature of 
the non-adiabatic H level crossing, independent of mass hierarchy, 
as we have seen in Sec.~\ref{nonadiab}. 
In the Garching flux model G1, the ratios of average neutrino energies 
and the widths of energy distributions of $\nu_e$ and $\nu_x$ to 
$\bar{\nu}_e$ are higher by about 7 - 20\% in the flux pattern $P_{21}$ 
than in $P_{22}$. At the same time, the luminosity ratios are higher in 
$P_{21}$ by about 20\%. 
In the model G2, the luminosity ratios are again higher by 30-40\% 
in the flux pattern $P_{21}$ than in $P_{22}$. 
The average energy and width ratios do not differ much except for 
$\frac{\langle \Delta E_{x} \rangle}{\langle \Delta E_{\bar{e}} \rangle}$
in which it is higher by about 30\% in $P_{21}$.

If the H level crossing is adiabatic the results depend upon 
the $\nu$ and $\bar{\nu}$ mass hierarchies. 
If they are both normal we are left with two flux patterns 
$P_{31}$ and $P_{32}$ (Sec.~\ref{adiab_NI}). 
In the Garching flux model G1, all the four ratios of average neutrino 
energies and their widths of the flux patterns 
$P_{31}$ are higher by about 10\% than those of $P_{32}$. 
The luminosity ratios also differ by about 20\%. 
In the model G2, average energy and width ratios do not 
show any appreciable difference, but the luminosity ratios 
are higher by 30-40\% in the flux pattern $P_{31}$ than in $P_{32}$. 
If the $\nu$ and $\bar{\nu}$ mass hierarchies are both inverted 
there is no way of signaling CPT violation in our method.

Therefore, it appears that there is a chance to uncover CPT 
violation within the accuracy which may be expected in a future detectors. 
Of course, we should not rely too much on the particular set of 
flux models to judge to what extent the different flux patterns are 
discriminable. 
But, the examination we have gone through in the above suggests 
that there are some possibilities of uncovering CPT violation 
by our method at least under circumstances helped by future 
terrestrial measurement.

\subsection{Supernova model dependence}

Here, we want to give a cautionary remark. 
Uncovering CPT violation along the way 
we discussed in this paper can only be claimed on the ground that 
the supernova simulations at the time of observation are reliable 
to certain extent. 
Even in the luckiest case in which we know  the mass hierarchy 
and that $\theta_{13}$ is in a region measurable by the next generation 
reactor and accelerator search, we need credibility in the flux model, 
roughly speaking, to 10-20\% level in the predictions to the ratios of 
average energy, width, and the luminosity of 
$\nu_e$ and $\nu_x$  to $\bar{\nu}_e$.

Suppose that a future measurement of supernova neutrino flux 
strongly suggest CPT violation by preferring one (or more) 
of the flux patterns which consists only of CPT violating mass patterns, 
or by disfavoring any of the CPT invariant patterns.
Then, one is tempted to conclude that CPT violation is signaled. 
The point is that the conclusion can be made firmer only by 
calibrating supernova simulation by accumulation of data gained 
by many explosions for consistency check. 
This point is a inherent weakness of the method of signaling CPT 
violation by supernova neutrino data.
We want to emphasize, however, that it is quite thinkable that we will 
have reliable model simulation of explosion in a timely way, 
by which neutrino flux can be predicted with high accuracies.
Neutrinos are the main engine of the explosion 
(they are like the pp neutrinos in the Sun) and, most probably, 
only the ordinary known physics is involved inside the core.

\subsection{Implications to analyses with CPT invariance}

As we mentioned at the end of Sec.~\ref{introduction}, part of the formulas 
and the analysis in this paper contain some informations useful for 
conventional analysis with CPT, in particular in the context of 
flavor-dependent reconstruction of three fluxes. 
We mention only a few points below, leaving full exposition to 
possible future works.

Look at the analytic formulas (\ref{F_31}) and  (\ref{F_23}) in 
Sec.~\ref{analytic}. 
From these equations, we notice the following features: 
In (\ref{F_31}) which applies to the case of normal mass hierarchy with 
adiabatic H resonance, 
the primary $\nu_{e}$ spectrum does not show up 
(or has a suppression factor of $s_{13}^2$) in the $\nu_{e}$ and 
$\bar{\nu}_{e}$ spectra observed at the terrestrial detectors.
It means that to reconstruct the primary $\nu_{e}$ spectrum 
accurate measurement of $\nu_{x}$ spectrum  together with 
those of $\nu_{e}$ and $\bar{\nu}_{e}$ is mandatory.
Similarly, the primary $\bar{\nu}_{e}$ spectrum does not appear 
(or has a suppression factor of $s_{13}^2$) in the $\nu_{e}$ and 
$\bar{\nu}_{e}$ spectra in (\ref{F_23}) which applies to the case 
of inverted mass hierarchy with adiabatic H resonance.

In these two cases it would be very difficult to reconstruct 
the primary $\nu_{e}$ ($\bar{\nu}_{e}$) spectrum in the case of 
normal (inverted) mass hierarchy with adiabatic H resonance, 
because spectral measurement of $\nu_{x}$ is difficult. 
(See, however, \cite{beacom} for a possible way out.)
We believe that it is one of the crucial problems in the program 
of flavor-dependent reconstruction of primary neutrino fluxes 
in the interior of supernova.

The case of normal and inverted mass hierarchies with 
non-adiabatic H resonance, (\ref{F_21}), may be the easiest one, 
relatively speaking,  
among the three CPT conserving patterns. 
It is because spectral measurement of $\bar{\nu}_{e}$ and 
$\nu_{e}$ (if possible) would allow us to determine all the three 
primary fluxes if the separation of two Fermi-Dirac type distributions 
with different temperatures is possible, as illustrated in \cite{MNTV}.

\section{Concluding remarks}
\label{conclusion}

We have discussed a method of using neutrinos from supernova to test 
CPT symmetry. 
Because of the possibility of having mass and mixing patterns 
of antineutrinos which can be different from neutrinos 32 
different scenarios are allowed which differ in neutrino mass patterns 
and (non-) adiabaticity of high-density resonance. 
They produce six different patterns of supernova neutrino energy 
spectra at the terrestrial neutrino detectors, apart from small modification 
due to earth matter effect. 
Among the six patterns, three of them contain only the CPT violating cases. 
Even in the mixed cases of CPT invariance and violation 
additional input on the values of $\theta_{13}$ from reactor and 
accelerator experiments further enhance the possibility of identifying 
CPT violation.

Future galactic supernovae watched by arrays of massive detectors 
may allow flavor-dependent reconstruction of three species of 
neutrino spectra, $\nu_e$, $\bar{\nu}_{e}$ and $\nu_x$ 
(a collective notation for $\nu_{\mu}$, $\bar{\nu}_{\mu}$, $\nu_{\tau}$, and 
$\bar{\nu}_{\tau}$). 
Assuming capability of obtaining such the informations, 
which are also required to diagnose supernova core in conventional 
analysis with CPT, we have shown that one of the three CPT violating 
patterns may be singled out observationally. 
The help by the other measurement of lepton mixing parameters by 
reactor and accelerator experiments would help to identify CPT violating 
cases. 
 Thus, we have shown that supernova neutrino can be a 
powerful tool to detect possible gross violation of CPT symmetry 
such as different mass patterns of neutrinos and antineutrinos. 
We emphasize the potential power of the method; 
It may allow to disentangle different (1-3) and (1-2) mass hierarchies 
both in neutrino and antineutrino sectors in a single ``bang''. 
%To obtain the same informations in the accelerator experiments, we would need projects of billions of dollars which need several decades to carry out.

However, we also noticed the weakness of the method.
We have just mentioned the flux model dependence in the previous section. 
Another drawback of the method concerns with its weakness at 
the precision test. 
In this paper, we have focused on the possibility of 
detecting CPT violation through identifying unequal mass patterns 
in neutrino and antineutrino sectors.
If the two CPT violation in masses as well as mixing angles coexist, 
however, it would be very difficult to 
clearly distinguish the six different patterns of neutrino flavor conversion, 
the topics we are unable to address in this paper.  
Therefore, it is important to have stringent bound on difference 
between neutrino and antineutrino mixing angles 
in order for the supernova method for testing CPT to work. 
Similarly, the analysis would be very complicated if $\theta_{13}$ 
is in the intermadiate region between adiabatic and non-adiabatic H resonance.

Suppose that CPT violation is signaled with supernova neutrinos 
in the way described in this paper. Then, one may feel that the 
confirmation by using man made neutrino beam 
necessary because of the fundamental importance of CPT symmetry.
Then, the question is; 
Is it possible to confirm CPT violation by the alternative methods?
Fortunately, the answer is yes. 
The possibility of using two detectors at the different baseline with 
$\nu$ beam only (not $\bar{\nu}$) to determine the neutrino mass hierarchy 
is proposed some time ago \cite{nu-nu}. 
Therefore, hypothesis of different mass hierarchies of neutrinos 
and antineutrinos can, in principle, be tested by separate measurement 
using the $\pi^+$ and $\pi^-$ beams. 
%For recent discussions about the possibility, see e.g., \cite{recent}.  
%
Determining the sign of $\Delta m^2_{21}$ in antineutrino sector 
is harder to carry out, as discussed in \cite{gouvea}. 
In any way, distinguishing neutrino mass hierarchy is a very 
challenging experiment and large-scale apparatus is required. 
We emphasize, therefore, that indication of CPT violation given 
by supernova neutrinos can give a good starting point of vital 
search for the totally unexpected phenomenon of 
CPT violation.

%%%%%%%%%%%%%%%% acknowledgments %%%%%%%%%%%%%
\begin{acknowledgments}
One of the authors (H.M.) thanks Hiroshi Nunokawa for discussions 
during a visit to 
Departamento de F\'{\i}sica, Pontif{\'\i}cia Universidade Cat{\'o}lica 
do Rio de Janeiro, where this work was completed. 
This work was supported in part by the Grant-in-Aid for Scientific Research, 
No. 16340078, Japan Society for the Promotion of Science.

\end{acknowledgments}
%%%%%%%%%%%%%%%%%%%%%%%%%%%%%%%%%%%%%%

%%%%%%%%%%%%%%%%%%%%%%%%%%%%%%


\begin{thebibliography}{99}


\bibitem {weinberg}
S.~Weinberg, {\it The Quantum Theory of Fields I} 
(Cambridge University Press, New York, 1995).


\bibitem{PDG}
S.~Eidelman {\it et al.}  [Particle Data Group Collaboration],
%``Review of particle physics,''
Phys.\ Lett.\ B {\bf 592}, 1 (2004).
%%CITATION = PHLTA,B592,1;%%


\bibitem{LSND}
A.~Aguilar {\it et al.}  [LSND Collaboration],
%``Evidence for neutrino oscillations from the observation of anti-nu/e
%appearance in a anti-nu/mu beam,''
Phys.\ Rev.\ D {\bf 64}, 112007 (2001)
[arXiv:hep-ex/0104049].
%%CITATION = HEP-EX 0104049;%%
See, however, 
B.~Armbruster {\it et al.}  [KARMEN Collaboration],
  %``Upper limits for neutrino oscillations anti-nu/mu $\to$ anti-nu/e from
  %muon decay at rest,''
  Phys.\ Rev.\ D {\bf 65}, 112001 (2002)
  [arXiv:hep-ex/0203021] 
  %%CITATION = HEP-EX 0203021;%%
for not completely consistent result. 


\bibitem{MY00}
H.~Murayama and T.~Yanagida,
%``LSND, SN1987A, and CPT violation,''
Phys.\ Lett.\ B {\bf 520}, 263 (2001)
[arXiv:hep-ph/0010178].
%%CITATION = HEP-PH 0010178;%%


\bibitem {SKatm}
Y.~Fukuda {\it et al.}  [Kamiokande Collaboration],
%``Atmospheric muon-neutrino / electron-neutrino ratio in the multiGeV energy range,''
Phys.\ Lett.\ B {\bf 335}, 237 (1994);
%%CITATION = PHLTA,B335,237;%%
%
Y.~Fukuda {\it et al.}  [Super-Kamiokande Collaboration],
%``Evidence for oscillation of atmospheric neutrinos,''
Phys.\ Rev.\ Lett.\  {\bf 81}, 1562 (1998)
[arXiv:hep-ex/9807003];
%%CITATION = HEP-EX 9807003;%%
Y.~Ashie {\it et al.}  [Super-Kamiokande Collaboration],
%``Evidence for an oscillatory signature in atmospheric neutrino oscillation,''
Phys.\ Rev.\ Lett.\  {\bf 93} (2004) 101801
[arXiv:hep-ex/0404034];
%%CITATION = HEP-EX 0404034;%%
Y.~Ashie {\it et al.}  [Super-Kamiokande Collaboration],
  %``A measurement of atmospheric neutrino oscillation parameters by
  %Super-Kamiokande I,''
  arXiv:hep-ex/0501064.
  %%CITATION = HEP-EX 0501064;%%


\bibitem {solar}
B.~T.~Cleveland {\it et al.},
%``Measurement Of The Solar Electron Neutrino Flux With The Homestake  Chlorine Detector,''
Astrophys.\ J.\  {\bf 496}, 505 (1998);
%%CITATION = ASJOA,496,505;%%
%
J.~N.~Abdurashitov {\it et al.}  [SAGE Collaboration],
%``Measurement of the solar neutrino capture rate with gallium metal,''
Phys.\ Rev.\ C {\bf 60}, 055801 (1999)
[arXiv:astro-ph/9907113];
%%CITATION = ASTRO-PH 9907113;%%
%
W.~Hampel {\it et al.}  [GALLEX Collaboration],
%``GALLEX solar neutrino observations: Results for GALLEX IV,''
Phys.\ Lett.\ B {\bf 447}, 127 (1999);
%%CITATION = PHLTA,B447,127;%%
%
S.~Fukuda {\it et al.}  [Super-Kamiokande Collaboration],
  %``Determination of solar neutrino oscillation parameters using 1496 days  of
  %Super-Kamiokande-I data,''
  Phys.\ Lett.\ B {\bf 539}, 179 (2002)
  [arXiv:hep-ex/0205075];
  %%CITATION = HEP-EX 0205075;%%
%
M.~B.~Smy {\it et al.}  [Super-Kamiokande Collaboration],
  %``Precise measurement of the solar neutrino day/night and seasonal  variation
  %in Super-Kamiokande-I,''
  Phys.\ Rev.\ D {\bf 69}, 011104 (2004)
  [arXiv:hep-ex/0309011];
  %%CITATION = HEP-EX 0309011;%%
%
Q.~R.~Ahmad {\it et al.}  [SNO Collaboration],
%``Measurement of the charged current interactions produced by B-8  solar neutrinos at the Sudbury Neutrino Observatory,''
Phys.\ Rev.\ Lett.\  {\bf 87}, 071301 (2001)
[arXiv:nucl-ex/0106015];
%%CITATION = NUCL-EX 0106015;%%
%Q.~R.~Ahmad {\it et al.}  [SNO Collaboration],
%``Direct evidence for neutrino flavor transformation from neutral-current  interactions in the Sudbury Neutrino Observatory,''
%Phys.\ Rev.\ Lett.\  
{\it ibid.} {\bf 89}, 011301 (2002)
[arXiv:nucl-ex/0204008]; 
%%CITATION = NUCL-EX 0204008;%%
B.~Aharmim {\it et al.}  [SNO Collaboration],
  %``Electron energy spectra, fluxes, and day-night asymmetries of B-8 solar
  %neutrinos from the 391-day salt phase SNO data set,''
  arXiv:nucl-ex/0502021.
  %%CITATION = NUCL-EX 0502021;%%


\bibitem{KamLAND}
K.~Eguchi {\it et al.} [KamLAND Collaboration],
Phys.\ Rev.\ Lett.\  {\bf 90}, 021802 (2003) 
[arXiv:hep-ex/0212021]; 
%%CITATION = HEP-EX 0212021;%%
T. ~Araki {\it et al.} [KamLAND Collaboration],
%``Measurement of neutrino oscillation with KamLAND: Evidence of spectral
%distortion,''
Phys.\ Rev.\ Lett.\  {\bf 94}, 081801 (2005) 
[arXiv:hep-ex/0406035].
%%CITATION = HEP-EX 0406035;%%


\bibitem {K2K}
M.~H.~Ahn {\it et al.}  [K2K Collaboration],
Phys.\ Rev.\ Lett.\  {\bf 90}, 041801 (2003) 
[arXiv:hep-ex/0212007];
%%CITATION = HEP-EX 0212007;%%
E.~Aliu {\it et al.}  [K2K Collaboration],
%``Evidence for muon neutrino oscillation in an accelerator-based experiment,''
Phys.\ Rev.\ Lett.\  {\bf 94}, 081802 (2005) 
[arXiv:hep-ex/0411038].
%%CITATION = HEP-EX 0411038;%%


\bibitem{Barenboim}
G.~Barenboim, L.~Borissov, J.~Lykken and A.~Y.~Smirnov,
%``Neutrinos as the messengers of CPT violation,''
JHEP {\bf 0210}, 001 (2002)
[arXiv:hep-ph/0108199];
%%CITATION = HEP-PH 0108199;%%
G.~Barenboim, L.~Borissov and J.~Lykken,
%``Neutrinos that violate CPT, and the experiments that love them,''
Phys.\ Lett.\ B {\bf 534}, 106 (2002)
[arXiv:hep-ph/0201080];
%%CITATION = HEP-PH 0201080;%%
G.~Barenboim, J.~F.~Beacom, L.~Borissov and B.~Kayser,
%``CPT violation and the nature of neutrinos,''
Phys.\ Lett.\ B {\bf 537}, 227 (2002)
[arXiv:hep-ph/0203261];
%%CITATION = HEP-PH 0203261;%%
G.~Barenboim and J.~Lykken,
%``A model of CPT violation for neutrinos,''
Phys.\ Lett.\ B {\bf 554}, 73 (2003)
[arXiv:hep-ph/0210411].
%%CITATION = HEP-PH 0210411;%%


\bibitem{Strumia02}
A.~Strumia,
%``Interpreting the LSND anomaly: Sterile neutrinos or CPT-violation  or...?,''
Phys.\ Lett.\ B {\bf 539}, 91 (2002)
[arXiv:hep-ph/0201134].
%%CITATION = HEP-PH 0201134;%%

\bibitem{Gouvea02}
A.~De Gouvea,
%``Can a CPT violating ether solve all electron (anti)neutrino puzzles?,''
Phys.\ Rev.\ D {\bf 66}, 076005 (2002)
[arXiv:hep-ph/0204077].
%%CITATION = HEP-PH 0204077;%%


\bibitem{concha03}
M.~C.~Gonzalez-Garcia, M.~Maltoni and T.~Schwetz,
%``Status of the CPT violating interpretations of the LSND signal,''
Phys.\ Rev.\ D {\bf 68}, 053007 (2003)
[arXiv:hep-ph/0306226].
%%CITATION = HEP-PH 0306226;%%


\bibitem{SK_CPT}
E.~Kearns, [for the Super-Kamiokande Collaboration],
Talk given at  {\it XXIst International Conference on Neutrino 
Physics and Astrophysics},  Paris, France, June 14-19.

%C.~Saji, Talk at the 5th Workshop on 
%``Neutrino Oscillations and their Origin'' (NOON2004), February 11-15, 2004, Odaiba, Tokyo, Japan, 
%http://www-sk.icrr.u-tokyo.ac.jp/noon2004/


\bibitem{Bahcall02}
J.~N.~Bahcall, V.~Barger and D.~Marfatia,
%``How accurately can one test CPT conservation with reactor and solar
%neutrino experiments?,''
Phys.\ Lett.\ B {\bf 534}, 120 (2002)
[arXiv:hep-ph/0201211].
%%CITATION = HEP-PH 0201211;%%

\bibitem{MNTZ}
H.~Minakata, H.~Nunokawa, W.~J.~C.~Teves and R.~Zukanovich Funchal, 
Phys.\ Rev.\ D  {\bf 71}  (2004) 013005 
[arXiv:hep-ph/0407326];
%%CITATION = HEP-PH 0407326;%%
%H.~Minakata, H.~Nunokawa, W.~J.~C.~Teves and R.~Z.~Funchal,
%``Reactor measurement of theta(12): Secret of the power,''
arXiv:hep-ph/0501250.
%%CITATION = HEP-PH 0501250;%%


\bibitem{Bilenky01}
S.~M.~Bilenky, M.~Freund, M.~Lindner, T.~Ohlsson and W.~Winter,
%``Tests of CPT invariance at neutrino factories,''
Phys.\ Rev.\ D {\bf 65}, 073024 (2002)
[arXiv:hep-ph/0112226].
%%CITATION = HEP-PH 0112226;%%


\bibitem{dighe-smi}
A.~S.~Dighe and A.~Y.~Smirnov,
%``Identifying the neutrino mass spectrum from the neutrino burst from a
%supernova,''
Phys.\ Rev.\ D {\bf 62}, 033007 (2000)
[arXiv:hep-ph/9907423].
%%CITATION = HEP-PH 9907423;%%
 

\bibitem{diagnostics}
See, e.g., 
J.~F.~Beacom, Talk at Neutrino Workshop at Institute for Nuclear Theory, Seattle, Washington, September 21-23, 2000; 
%
H.~Minakata, 
%``Diagnostics of Supernova Neutrinos by Superkamiokande"
Talk at Frontiers in Particle Astrophysics and Cosmology; 
EuroConference on Neutrinos in the Universe, Lenggries, Germany, September 29 - October 4, 2001. 


\bibitem{MN_inverted} 
H.~Minakata and H.~Nunokawa,
  %``Inverted hierarchy of neutrino masses disfavored by supernova 1987A,''
  Phys.\ Lett.\ B {\bf 504}, 301 (2001)
  [arXiv:hep-ph/0010240].
  %%CITATION = HEP-PH 0010240;%%


\bibitem{SSB94}
  A.~Y.~Smirnov, D.~N.~Spergel and J.~N.~Bahcall,
  %``Is large lepton mixing excluded?,''
  Phys.\ Rev.\ D {\bf 49}, 1389 (1994)
  [arXiv:hep-ph/9305204].
  %%CITATION = HEP-PH 9305204;%%

\bibitem{jegerlehner}
  B.~Jegerlehner, F.~Neubig and G.~Raffelt,
  %``Neutrino Oscillations and the Supernova 1987A Signal,''
  Phys.\ Rev.\ D {\bf 54}, 1194 (1996)
  [arXiv:astro-ph/9601111].
  %%CITATION = ASTRO-PH 9601111;%%

\bibitem{barger02}
  V.~Barger, D.~Marfatia and B.~P.~Wood,
  %``Supernova 1987A did not test the neutrino mass hierarchy,''
  Phys.\ Lett.\ B {\bf 532}, 19 (2002)
  [arXiv:hep-ph/0202158].
  %%CITATION = HEP-PH 0202158;%%


\bibitem{raffelt}
G.~Raffelt, private communications in 2002.
 


\bibitem {MSW}
S.~P.~ Mikheyev and A.~Yu.~ Smirnov,
Yad.\ Fiz.\ {\bf 42}, 1441 (1985)
[ Sov.\ J. Nucl.\ Phys.\ {\bf 42}, 913 (1985)];
Nuovo Cim.\ C {\bf 9}, 17 (1986); 
%%CITATION = NUCIA,C9,17;%%
L.~Wolfenstein,
Phys.\ Rev.\ D {\bf 17}, 2369 (1978).
%%CITATION = PHRVA,D17,2369;%%


\bibitem {MNS}
Z.~Maki, M.~Nakagawa and S.~Sakata,
%``Remarks On The Unified Model Of Elementary Particles,''
Prog.\ Theor.\ Phys.\  {\bf 28}, 870 (1962).
%%CITATION = PTPKA,28,870;%%
See also, B.~Pontecorvo, 
Zh. Eksp. Teor. Fyz. {\bf 53}, 1717 (1967) 
[Sov. Phys. JETP {\bf 26}, 984 (1968)].  


\bibitem{MN90}
 H.~Minakata and H.~Nunokawa,
%``Light Neutrinos As Cosmological Dark Matter And The Next Supernova,''
Phys.\ Rev.\ D {\bf 41}, 2976 (1990).
%%CITATION = PHRVA,D41,2976;%%


\bibitem{gouvea}
  A.~de Gouvea and C.~Pena-Garay,
  %``Probing new physics by comparing solar and KamLAND data,''
  Phys.\ Rev.\ D {\bf 71}, 093002 (2005)
  [arXiv:hep-ph/0406301].
  %%CITATION = HEP-PH 0406301;%%


\bibitem{JPARC}
Y.~Itow {\it et al.}, arXiv:hep-ex/0106019. 
%%CITATION = HEP-EX 0106019;%%
An updated version: \\
http://neutrino.kek.jp/jhfnu/loi/loi.v2.030528.pdf 


\bibitem {NOVA}
D.~Ayres {\it et al.}  [Nova Collaboration],
  arXiv:hep-ex/0503053. 
  %%CITATION = HEP-EX 0503053;%%


\bibitem {SPL}
J.~J.~Gomez-Cadenas {\it et al.}, 
%[CERN working group on Super Beams Collaboration] 
arXiv:hep-ph/0105297.
%%CITATION = HEP-PH 0105297;%%


\bibitem {beta}
J.~Burguet-Castell, D.~Casper, E.~Couce, J.~J.~Gomez-Cadenas and P.~Hernandez,
  %``Optimal beta-beam at the CERN-SPS,''
  arXiv:hep-ph/0503021.
  %%CITATION = HEP-PH 0503021;%%


\bibitem{reactor} 
See, for example, 
H.~Minakata, H.~Sugiyama, O.~Yasuda, K.~Inoue, and F.~Suekane, 
Phys.\ Rev.\ D {\bf 68}, 033017 (2003)
[arXiv:hep-ph/0211111]; 
%%CITATION = HEP-PH 0211111;%%
K.~Anderson  {\it et al.},
White Paper Report on Using Nuclear Reactors to Search for a 
Value of $\theta_{13}$, arXiv:hep-ex/0402041. 
%%CITATION = HEP-EX 0402041;%% 


\bibitem{nufact}
C.~Albright {\it et al.},
%``Physics at a neutrino factory,''
arXiv:hep-ex/0008064; 
%%CITATION = HEP-EX 0008064;%%
M.~Apollonio {\it et al.}, arXiv:hep-ph/0210192.
%%CITATION = HEP-PH 0210192;%%


 \bibitem{recent}
% For recent discussions, see e.g., 
    M.~Ishitsuka, T.~Kajita, H.~Minakata and H.~Nunokawa,
  %``Resolving neutrino mass hierarchy and CP degeneracy by two identical
  %detectors with different baselines,''
  Phys.\ Rev.\ D {\bf 72}, 033003 (2005)
  [arXiv:hep-ph/0504026]; 
  %%CITATION = HEP-PH 0504026;%%
  O.~Mena Requejo, S.~Palomares-Ruiz and S.~Pascoli,
  %``Super-NOvA: A long-baseline neutrino experiment with two off-axis
  %detectors,''
  arXiv:hep-ph/0504015;
  %%CITATION = HEP-PH 0504015;%%
K.~Hagiwara, N.~Okamura and K.~i.~Senda,
  %``Solving the neutrino parameter degeneracy by measuring the T2K off-axis
  %beam in Korea,''
  arXiv:hep-ph/0504061.
  %%CITATION = HEP-PH 0504061;%%


\bibitem{nakahata}
M.~Nakahata,
  %``Future solar neutrino experiments,''
  Nucl.\ Phys.\ Proc.\ Suppl.\  {\bf 145}, 23 (2005).
  %%CITATION = NUPHZ,145,23;%%
 
 
\bibitem{goswami}
A.~Bandyopadhyay, S.~Choubey, S.~Goswami and S.~T.~Petcov,
  %``High precision measurements of Theta(solar) in solar and reactor  neutrino
  %experiments,''
  arXiv:hep-ph/0410283.
  %%CITATION = HEP-PH 0410283;%%
  

  \bibitem{livermore}
T.~Totani, K.~Sato, H.~E.~Dalhed and J.~R.~Wilson,
%``Future detection of supernova neutrino burst and explosion mechanism,''
Astrophys.\ J.\  {\bf 496}, 216 (1998)
[astro-ph/9710203].
%%CITATION = ASTRO-PH 9710203;%%



\bibitem{MNTV}
  H.~Minakata, H.~Nunokawa, R.~Tomas and J.~W.~F.~Valle,
  %``Probing supernova physics with neutrino oscillations,''
  Phys.\ Lett.\ B {\bf 542}, 239 (2002)
  [arXiv:hep-ph/0112160].
  %%CITATION = HEP-PH 0112160;%%


\bibitem{luna-smi2}
  C.~Lunardini and A.~Y.~Smirnov,
  %``Probing the neutrino mass hierarchy and the 13-mixing with supernovae,''
  JCAP {\bf 0306}, 009 (2003)
  [arXiv:hep-ph/0302033].
  %%CITATION = HEP-PH 0302033;%%


\bibitem{MKeil}
M.~T.~Keil,
``Supernova neutrino spectra and applications to flavor oscillations,''
PhD thesis TU M\"unchen 2003
[astro-ph/0308228].
%%CITATION = ASTRO-PH 0308228;%%


\bibitem{Keil02}
M.~T.~Keil, G.~G.~Raffelt and H.~T.~Janka,
%``Monte Carlo study of supernova neutrino spectra formation,''
Astrophys.\ J.\  {\bf 590} (2003) 971
[astro-ph/0208035];
%%CITATION = ASTRO-PH 0208035;%%
G.~G.~Raffelt, M.~T.~Keil, R.~Buras, H.~T.~Janka and M.~Rampp,
  %``Supernova neutrinos: Flavor-dependent fluxes and spectra,''
  arXiv:astro-ph/0303226.
  %%CITATION = ASTRO-PH 0303226;%%


  \bibitem{garching}
R.~Buras, H.~T.~Janka, M.~T.~Keil, G.~G.~Raffelt and M.~Rampp,
%``Electron-neutrino pair annihilation: A new source for muon and tau neutrinos in supernovae,'' 
Astrophys.\ J.\  {\bf 587}, 320 (2003)
[astro-ph/0205006].
%%CITATION = ASTRO-PH 0205006;%%


\bibitem{tomas_etal}
  R.~Tomas, M.~Kachelriess, G.~Raffelt, A.~Dighe, H.~T.~Janka and L.~Scheck,
  %``Neutrino signatures of supernova shock and reverse shock propagation,''
  arXiv:astro-ph/0407132.
  %%CITATION = ASTRO-PH 0407132;%%
  
 
\bibitem{dutta}
G.~Dutta, D.~Indumathi, M.~V.~N.~Murthy and G.~Rajasekaran,
  %``Neutrinos from stellar collapse: Effects of flavour mixing,''
  Phys.\ Rev.\ D {\bf 61}, 013009 (2000)
  [arXiv:hep-ph/9907372].
  %%CITATION = HEP-PH 9907372;%%
  %
G.~Dutta, D.~Indumathi, M.~V.~N.~Murthy and G.~Rajasekaran,
  %``Neutrinos from stellar collapse: Comparison of signatures in water and
  %heavy water detectors,''
  Phys.\ Rev.\ D {\bf 64}, 073011 (2001)
  [arXiv:hep-ph/0101093].
  %%CITATION = HEP-PH 0101093;%%


\bibitem{todai}
K.~Takahashi, M.~Watanabe, K.~Sato and T.~Totani,
  %``Effects of neutrino oscillation on the supernova neutrino spectrum,''
  Phys.\ Rev.\ D {\bf 64}, 093004 (2001)
  [arXiv:hep-ph/0105204];
  %%CITATION = HEP-PH 0105204;%%
%
 K.~Takahashi and K.~Sato,
  %``Effects of neutrino oscillation on supernova neutrino: Inverted mass
  %hierarchy,''
  Prog.\ Theor.\ Phys.\  {\bf 109}, 919 (2003)
  [arXiv:hep-ph/0205070].
  %%CITATION = HEP-PH 0205070;%%


\bibitem{barger}
  V.~Barger, D.~Marfatia and B.~P.~Wood,
  %``Inverting a supernova: Neutrino mixing, temperatures and binding  energy,''
  Phys.\ Lett.\ B {\bf 547}, 37 (2002)
  [arXiv:hep-ph/0112125].
  %%CITATION = HEP-PH 0112125;%%

 
\bibitem{luna-smi1}
C.~Lunardini and A.~Y.~Smirnov,
  %``Supernova neutrinos: Earth matter effects and neutrino mass spectrum,''
  Nucl.\ Phys.\ B {\bf 616}, 307 (2001)
  [arXiv:hep-ph/0106149].
  %%CITATION = HEP-PH 0106149;%%


 \bibitem{nakamura}
K.~Nakamura, Talk at Next Generation of Nucleon Decay and 
Neutrino Detectors (NNN05), 
Aussois, Savoie, France, April 7-9, 2005. 

 
 \bibitem{beacom}
  J.~F.~Beacom, W.~M.~Farr and P.~Vogel,
  %``Detection of supernova neutrinos by neutrino proton elastic scattering,''
  Phys.\ Rev.\ D {\bf 66}, 033001 (2002)
  [arXiv:hep-ph/0205220].
  %%CITATION = HEP-PH 0205220;%%
  
 
\bibitem{nu-nu}
  H.~Minakata, H.~Nunokawa and S.~J.~Parke,
  %``The complementarity of eastern and western hemisphere long-baseline
  %neutrino oscillation experiments,''
  Phys.\ Rev.\ D {\bf 68}, 013010 (2003)
  [arXiv:hep-ph/0301210]; 
  %%CITATION = HEP-PH 0301210;%%
  P.~Huber, M.~Lindner and W.~Winter,
  %``Synergies between the first-generation JHF-SK and NuMI superbeam
  %experiments,''
  Nucl.\ Phys.\ B {\bf 654}, 3 (2003)
  [arXiv:hep-ph/0211300].
  %%CITATION = HEP-PH 0211300;%%
  
 

\end{thebibliography}
\end{document}